\documentclass[preprint,authoryear,12pt]{elsarticle}
\usepackage{amssymb,color}

\journal{arxiv}

\newcommand{\beq}{\begin{equation}}
\newcommand{\beql}[1]{\begin{equation}\label{#1}}
\newcommand{\eeq}{\end{equation}}
\newcommand{\bea}{\begin{eqnarray}}
\newcommand{\eea}{\end{eqnarray}}
\newcommand{\lamd}{\dot{\lambda}}
\newcommand{\lamp}{\lambda_p}
\newcommand{\lamg}{\lambda_g}
\newcommand{\kap}{\kappa}
\newcommand{\kapd}{\dot{\kappa}}

\newcommand{\kapt}{\kappa_n}

\newcommand{\eps}{\varepsilon}
\newcommand{\epse}{\varepsilon_e}
\newcommand{\epsp}{\varepsilon_p}
\newcommand{\epsdp}{\dot\varepsilon_p}
\newcommand{\sigY}{\sigma_Y}
\newcommand{\sigc}{\sigma_c}
\renewcommand{\lg}{l_g} 
\newcommand{\half}{\mbox{$\frac{1}{2}$}}
\newcommand{\Lp}{L_p}
\newcommand{\Ip}{{\cal I}_p}

\newcommand{\ignore}[1]{}
\newcommand{\dx}{\;\mathrm{d}x}
\newcommand{\dxi}{\;\mathrm{d}\xi}
\newcommand{\sumdL}{\sum_{\partial{\cal L}}}
\newcommand{\intL}{\int_{\cal L}}

\newcommand{\eee}{{\rm e}}
\newcommand{\sgn}{{\rm sgn}}
\newcommand{\sig}{\sigma}
\newcommand{\erf}{\mbox{erf}}
\newcommand{\newstuff}[1]{ #1}

\begin{document}

\begin{frontmatter}

\title{Localization Analysis of Variationally Based Gradient Plasticity Model}

\author[label1]{Milan Jir\'{a}sek}
\author[label1]{Ond\v{r}ej  Roko\v{s}}
\author[label1]{Jan Zeman}

\address[label1]{Department of Mechanics, 
Faculty of Civil Engineering, Czech Technical University in Prague}

\begin{abstract}
The paper presents analytical or semi-analytical solutions for the
formation and evolution of localized plastic zone in a uniaxially loaded
bar with variable cross-sectional area. A variationally based formulation
of explicit gradient plasticity with linear softening is used, and the
ensuing jump conditions and boundary conditions are discussed.
Three cases with different regularity of the stress distribution are
considered, and the problem is converted to a dimensionless form. 
Relations linking the load level, size of the plastic zone,
distribution of plastic strain and plastic elongation of the bar
are derived and compared to another, previously analyzed gradient formulation.
\end{abstract}


\end{frontmatter}

\section{One-Dimensional Softening Plasticity Model}
\label{subsec:1.2}

For many materials, the stress-strain diagrams characterizing their
mechanical behavior exhibit the so-called softening branches, with
decreasing stress at increasing strain (and thus with a negative
tangent stiffness). The physical origin of this intriguing phenomenon
is in the initiation, propagation and coalescence of defects such as
microcracks or microvoids. Softening-induced localization of inelastic
processes into narrow zones often acts as a precursor to failure.
Proper modeling of the entire failure process requires an objective
description of the localized process zone and its evolution.

Perhaps the most popular class of inelastic material models is 
represented by the theory of (elasto-)plasticity. The present paper focuses
on the localization properties of softening plasticity models.
To allow for analytical solutions, all considerations are done
in the one-dimensional context, referring to the case of a straight
bar under uniaxial loading as the typical paradigm. However, the analysis
is nontrivial due to the fact that a variable cross-sectional area
is considered, and a regularized formulation of softening plasticity
is used.

\subsection{Classical Formulation}

In the small-strain range, classical elastoplasticity is based 
on the additive split of the total strain into the elastic part
and the plastic part. The elastic strain is linked to the stress
by Hooke's law, while the plastic strain can grow only if the stress
level attains the yield limit, which is mathematically indicated by
zero value of the yield function. The oriented direction of the plastic
strain rate is specified by the flow rule and the evolution of the
yield surface (set of plastic stress states in the stress space)  is
described by the hardening/softening law. For simplicity, we assume
linear softening, i.e., linear dependence of the current yield stress
on the cumulative plastic strain. 
Description
of the stress-strain relation by a bilinear diagram is certainly a rough approximation,
but it can reflect the main features of elastoplasticity with 
softening and serve as a prototype model, for which analytical solutions exist.

In the one-dimensional setting, the basic equation can be summarized 
as follows:
\bea
\label{eqgp1}
\sig &=& E\epse = E(\eps-\epsp)
\\
\label{eq-gp2}
f(\sig,\kappa) &=& \vert \sig \vert - \sigY(\kappa)
\\
\label{eqgp2a}
\sigY(\kappa)&=&\sig_0+H\kappa
\\
\label{eq-gp3}
\epsdp&=&\lamd\,\sgn\sig
\\
\label{eqgp5}
\kapd&=&\vert\epsdp\vert
\\
\label{eqgp4}
\lamd\ge 0, \hskip 10mm f(\sig,\kappa)&\le& 0, \hskip 10 mm \lamd f(\sig,\kappa)=0
\eea
Here, $\sig$ is the stress, $\eps$ is the (total) strain, $\epse$ and $\epsp$
are its elastic and plastic parts, $E$ is the elastic modulus, $f$ is the yield function,
$\sigY$ is the current yield stress,
$\sig_0$ is the initial yield stress, $H<0$ is the softening modulus,
$\lambda$ is the plastic multiplier and  $\kappa$ is 
the cumulative plastic strain.
The overdot denotes differentiation 
with respect to time. A more detailed discussion of this specific
problem is available in \cite{jzv10} and a broad background of the theory
of plasticity e.g.\ in \cite{Lub90} or \cite{JirBaz01}.

If we restrict attention to tensile loading (with possible elastic unloading,
but never with a reversal of plastic flow), then the plastic strain $\epsp$, cumulative
plastic strain $\kappa$ and plastic multiplier $\lambda$ are all equal.
We will use $\kappa$ as the primary symbol for (cumulative) plastic strain 
and rewrite equations (\ref{eqgp1}) and  (\ref{eqgp4}) as
\beql{eqgp6}
\sig = E(\eps-\kappa)
\eeq
\beql{eqgp7x}
\kapd\ge 0, \hskip 10mm f(\sig,\kappa)\le 0, \hskip 10 mm \kapd f(\sig,\kappa)=0
\eeq

The above equations refer to uniaxial tension,
but formally the same framework can be used for the one-dimensional 
description of a shear problem. Normal stress and strain are then replaced 
by shear stress and strain, Young's modulus $E$ by the shear modulus $G$,
and the tensile yield stress by the shear yield stress.

\subsection{Standard Gradient Formulation}

It is well known that softening is a destabilizing factor that may lead
to localization of dissipative processes (in our case of plastic yielding)
into narrow zones. For classical continuum formulations with local dependence
between stress and strain, the thickness of such localized process zones is   
undetermined and may become arbitrarily small. The undesired consequence 
is that the structural response becomes excessively brittle and numerical
simulations suffer by pathological sensitivity to the discretization
parameters such as the size of elements used by the finite element method.
This has to be avoided, e.g.\ by introducing a regularization technique 
which enforces a nonzero
thickness of the localized process zone and thus nonvanishing dissipation
during the failure process.

In the one-dimensional setting, negative plastic modulus $H$ always leads
to localization of plastic strain. 
Consider a straight bar with perfectly uniform
properties, subjected to uniaxial tension (induced by applied displacement at one bar end). The response
remains uniform in the elastic range and also during plastic yielding with a positive plastic
modulus. For a negative (or vanishing) plastic modulus, uniqueness of the solution is 
lost right at the onset
of softening (or of perfectly plastic yielding). 
Stress distribution along the bar must still remain uniform due to the static
equilibrium conditions (in the absence of body forces), but a given stress level can
be attained by softening with increasing plastic strain, or by elastic unloading with no
plastic strain evolution. Which cross sections unload
and which exhibit softening remains completely arbitrary, 
and there is no lower bound on the total length of the
softening region(s). Therefore, infinitely many solutions exist, including solutions
with plastic strain evolution localized into extremely small regions. 
Even if the nonuniqueness of the solution is removed by a slight perturbation of the perfect
uniformity of the bar, the problem with localization of softening into arbitrarily small
regions (in fact into the weakest cross section) still persists.

Commonly used regularization techniques overcome the problem by suitable
enrichments of the governing equations. Such enrichments typically introduce
at least one additional parameter with the dimension of length (or a parameter
which can be combined with the traditional ones such that the result has the
dimension of length). This parameter reflects the intrinsic length scale of
the material and is related to the size and spacing of major heterogeneities
in the microstructure. The size of the process zone is then controled by the
choice of the length scale parameter.

In principle it is possible to construct regularized 
models with enriched kinematic and
equilibrium equations, e.g.\ strain-gradient plasticity or Cosserat-type 
models.
From the practical point of view it is more convenient to limit the  
enrichments to the constitutive equations describing the material
behavior and to keep the kinematic and equilibrium equations unchanged.
This class of approaches is usually referred to as nonlocal continuum 
theories in the broad sense. Nonlocality of the stress-strain relation
can be introduced by weighted spatial averaging of suitably chosen
internal variables, or by incorporation of gradients of such variables
into the constitutive description. Here we focus on the latter case,
in particular on its typical representative---the second-order
explicit gradient model that evolved from the work of Aifantis and
colleagues \cite{Aif84}.

The explicit gradient formulation of elastoplasticity is based on
incorporation of a term proportional to the Laplacean of cumulative
plastic strain into the softening law (\ref{eqgp2a}). In the one-dimensional
setting, the Laplacean reduces to the second derivative and the enriched
softening law reads
\beq
\label{eqgp2ax}
\sigY(\kappa)=\sig_0+H(\kappa+l^2\kappa'')
\eeq 
where $l$ is a new parameter with the dimension of length.

In a bar with perfectly uniform properties (cross section, yield stress,
softening modulus, etc.) and in the absence of body forces and inertia
forces, the stress is constant along the bar. The plastic zone, $\Ip$, is 
characterized by growing plastic strain ($\dot{\kappa}>0$) and vanishing
value of the yield function ($f=0$). Since the yield function is given
by (\ref{eq-gp2}), we conclude that the yield stress must be constant inside the
plastic zone, and then (\ref{eqgp2ax}) leads to a second-order differential
equation with constant coefficients and a constant right-hand side:  
\beql{eqgp123}
l^2\kappa''(x)+\kappa(x) = \frac{\sig-\sig_0}{H}
\eeq
As shown e.g.\ in \cite{bormuhl92}, the (most localized) plastic zone is an interval
of length $2\pi l$, arbitrarily placed along the bar.

\newstuff
{
Analytical solutions for several types of bars with variable cross sections
were presented in \cite{jzv10}. The governing equation 
\beql{eqgp124}
l^2A(x)\kappa''(x)+A(x)\kappa(x) = \frac{F-\sig_0A(x)}{H}
\eeq
was constructed from (\ref{eqgp123})
by  setting $\sigma(x)=F/A(x)$, where $A$ is a function describing the 
distribution of the cross-sectional area along the bar, 
and $F$ is the axial force transmitted by the bar, which is constant
(independent of $x$) because of static equilibrium.
}

In the present paper, we will use a modified formulation of the one-dimensional
gradient plasticity model, constructed by a variational approach.
Analytical or semi-analytical solutions will be derived and compared to
the results for the standard gradient formulation based on (\ref{eqgp124}).  
An important advantage of the variational formulations is that it permits
a consistent treatment of problems with discontinuous data, e.g.\ with
a jump in the cross-sectional area (leading to a jump in the stress field).

\subsection{Variational Gradient Formulation}
\label{sec:varfor}

The variational formulation of the second-order explicit gradient plasticity model considered here is inspired by the work of 
\cite{muhlaif91,Val96,sved96,SveRun97,SveRun98a,PolBorFus98,BorFusPol99,liestei01}.
In the one-dimensional case, it is derived from the functional
\bea\nonumber
\Pi(u,\kap) &=& \intL \half EA(u'-\kap)^2\dx + \intL\half HA\left(\kap^2-l^2\kap'^2\right)\dx 
+
\\
&&+\intL A\sig_0\kap\dx - \intL Abu\dx
\label{varfor1}
\eea
in which ${\cal L}$ denotes the interval that represents the entire bar, $A$ is the cross sectional area
and $b$ is the prescribed body force density in the longitudinal direction (per unit volume), introduced just for the sake
of generality but later set equal to zero. The first integral in (\ref{varfor1}) can be interpreted as the elastic
strain energy, the second as the plastic part of free energy, the third as the dissipated energy and the fourth
as the potential energy of external forces.

\newstuff
{
Functional $\Pi$ is considered in the space of all sufficiently smooth
displacement
fields $u$ that  satisfy the geometric (essential)
boundary conditions on the boundary $\partial{\cal L}$, 
and all sufficiently smooth and nonnegative plastic strain fields $\kap$.
}
In formal mathematical language, the domain of definition of functional $\Pi$ is the space
$V=V_u\times V_{\kap}$ where
\bea\label{varfor2}
V_u &=& \left\{u\in H_1({\cal L}) \;\;\vert\;\; u=\bar u \mbox{ on } \partial{\cal L} \mbox{ in the sense of traces}\right\} 
\\
V_\kap &=& \left\{\kap\in H_1({\cal L}) \;\;\vert\;\; \kap\ge 0 \mbox{ almost everywhere}\right\}
\label{varfor3}
\eea
This means that the functions describing the displacement and the plastic strain 
must be square-integrable and possess square-integrable generalized first derivatives,
but continuity of the first derivatives and existence of the second derivatives are not apriori required.

\newstuff
{
Due to the lack of convexity, the analysis can hardly rely on global
minimization of functional $\Pi$. Nevertheless, it is reasonable to expect
that stable solutions of the problem are associated with local minima
of $\Pi$. The subsequent derivations will be based on necessary conditions
of a local minimum, in particular, on nonnegative values of the first variation
(Fr\'{e}chet derivative) of functional $\Pi$ corresponding to all admissible
variations of fields $u$ and $\kap$. It will be demonstrated that such
an approach leads to a consistent set of conditions that describe the
problem and include the equilibrium equation, the complementarity conditions
governing the plastic flow, as well as appropriate boundary conditions at the 
physical boundary and regularity conditions at the elasto-plastic interfaces.
A complete analysis should also pay attention to the second variation,
which is related to stability issues. Analytical conditions for 
a non-negative second variation, derived for the simplest case of a bar
with uniform properties, are presented in Appendix~A.  

Strictly speaking, the variational approach should be applied in 
an incremental fashion, as discussed e.g.\ by \cite{Pet03}. However, 
for the present purpose it is fully sufficient to consider a total
formulation. It turns out that, for one-dimensional problems 
with expanding or stationary
plastic zones, the parameterized solutions constructed
in this way do not violate the irreversibility constraints and thus
represent physically admissible responses to given loading scenarios.
}

The first variation  of functional $\Pi$ defined in (\ref{varfor1})
can be expressed as
\bea\nonumber
\delta\Pi(\delta u,\delta\kap;u,\kap) &=& \intL EA(u'-\kap)(\delta u'-\delta\kap)\dx + 
\\ \nonumber
&&+\intL HA\left(\kap\delta\kap-l^2\kap'\delta\kap'\right)\dx +
\\
&&+\intL A\sig_0\delta\kap\dx - \intL Ab\delta u\dx
\label{varfor4}
\eea
where $\delta u$ is the displacement variation (difference between two admissible displacement fields taken
from $V_u$) and $\delta\kap$ is the variation of plastic strain (difference between two admissible plastic strain 
fields taken from $V_\kap$).  
Integration by parts of the terms with $\delta u'$ and $\delta\kap'$ leads to
\bea\nonumber
\delta\Pi(\delta u,\delta\kap;u,\kap) &=& 
- \intL \left[(EA(u'-\kap))'+Ab\right]\delta u\;\dx
\\&&\nonumber
+ \intL \left[HA\kap+(HAl^2\kap')'+A\sig_0-EA(u'-\kap)\right]\delta\kap\;\dx 
\\&&\nonumber
+\sumdL EA(u'-\kap)n\delta u- \sum_i[[EA(u'-\kap)\delta u]]_{x_i} 
\\&&
-\sumdL HAl^2\kap'n\delta\kap + \sum_i[[HAl^2\kap'\delta\kap]]_{x_i} 
\label{eqgp104}
\eea
where $\sumdL$ stands for the boundary integral, in the one-dimensional setting 
reduced to the sum over two points at the boundary of the interval ${\cal L}$, $n$ is the ``unit normal'',
equal to $-1$ at the left boundary and to $1$ at the right boundary, and the sums with subscript $i$ are taken over
all points at which the quantity in the double square brackets has a discontinuity. The double brackets denote the
jump of the quantity inside the brackets, defined as 
\beq
[[f]]_{x_i}=\lim_{x\rightarrow x_i^+} f(x) - \lim_{x\rightarrow x_i^-} f(x)
\eeq
As already explained,
the first variation $\delta\Pi$ must be nonnegative for all admissible variations $\delta u$ and $\delta\kap$.
By admissible variations we mean arbitrary changes of $u$ and $\kap$ for which $u+\delta u\in V_u$ and
$\kap+\delta\kap\in V_{\kap}$. 

Variations $\delta u$ are arbitrary inside ${\cal L}$ but vanishing on the boundary.
The expression multiplying $\delta u$ in the integral on the first line of (\ref{eqgp104}) must be identically equal to zero,
which provides the equilibrium conditions
\beq\label{eqgp106x}
(EA(u'-\kap))'+Ab=0
\eeq
Of course, $u'$ corresponds to the strain, $u'-\kap$ is the elastic strain, $E(u'-\kap)$ is the stress
and $EA(u'-\kap)$ is recognized as the axial force.
Since $\delta u$ is arbitrary inside ${\cal L}$, the jumps of $EA(u'-\kap)$ must vanish, i.e., the axial force
(not the stress) must
remain continuous.

The variation of $\kap$ is not completely arbitrary, because $\kap+\delta\kap$ must remain nonnegative.
If we define the plastic zone $\Ip=\{x\in{\cal L}\;\vert\;\kap(x)>0\}$, $\delta\kap$ can have an arbitrary 
sign inside $\Ip$ but must be nonnegative outside $\Ip$. Therefore, the expression multiplying $\delta\kap$
in the integral on the second line of (\ref{eqgp104}) must vanish in $\Ip$ but outside $\Ip$ it is only constrained
to be nonnegative. We recognize the resulting equation 
\beql{eq:50}
HA\kap+(HAl^2\kap')'+A\sig_0=EA(u'-\kap) \mbox{ for } x\in\Ip
\eeq
as the yield condition, and the resulting inequality
\beql{eq:50x}
HA\kap+(HAl^2\kap')'+A\sig_0\ge EA(u'-\kap) \mbox{ for } x\in{\cal L}\setminus\Ip 
\eeq
as the plastic admissibility condition.
The advantage of the variational formulation is that the cases of variable sectional area $A$,
softening modulus $H$ or internal length $l$ are covered in a systematic way, even in cases when
some of these quantities exhibit discontinuities.
From the last two lines of (\ref{eqgp104}) 
we obtain the corresponding jump conditions and also the boundary conditions.

On the physical boundary $\partial{\cal L}$, we get $HAl^2\kap'n= 0$ if the boundary point
belongs to $\Ip$, or  $HAl^2\kap'n\le 0$ if this point does not belong to $\Ip$. The first condition
means that if the plastic zone contains a point of the physical boundary, the homogeneous Neumann condition $\kap'=0$
should be imposed at that point. The second condition means that if the boundary point remains elastic ($\kap=0$ at the
boundary), the spatial derivative of plastic strain could, in principle, be nonzero. Since $H<0$ 
and $Al^2>0$, at the right end of the bar ($n=1$) the derivative $\kap'$ must not be negative. However,
if $\kap=0$ at the right boundary and $\kap\ge 0$ everywhere, $\kap'$ cannot be strictly positive at the
boundary and its only admissible value is again zero.

The jump conditions imply that if the quantity $HAl^2\kap'\delta\kap$ is discontinuous at some
point, its jump should be nonnegative for any admissible $\delta\kap$. 
\begin{itemize}
\item
For points inside  the plastic zone $\Ip$,
$\delta\kap$ can be positive as well as negative, and therefore $HAl^2\kap'$ must remain continuous. So in general we should
not enforce continuous differentiability of plastic strain, but rather continuity of the product 
$HAl^2\kap'$. This is important when the spatial distribution of the softening modulus or of the sectional
area is discontinuous. 
\item
For points outside the plastic zone, the variation $\delta\kap$
is nonnegative and so the jump of $HAl^2\kap'$  is in principle admissible but only if it is positive.
Inside the elastic zone, $\kap'$ vanishes and $HAl^2\kap'$ has no jump at all. However, at the elasto-plastic interface
(which is located at the boundary
of the plastic zone), the derivative of plastic strain could exhibit a jump from zero value in the elastic zone
to nonzero value in the plastic zone. At the left boundary of the plastic zone, the jump is equal to 
the value in the plastic zone.
Since $H$ is negative and $Al^2$ is positive, the limit of $\kap'$ (as we approach the boundary of the plastic zone
from inside) is allowed to be negative (or zero). However, $\kap'<0$ would mean that $\kap<0$ at some point inside
the plastic zone (because $\kap=0$ at the boundary of that zone), which is not admissible. Similar arguments can
be applied at the right boundary of the plastic zone, where the jump is minus the value in the plastic zone,
and therefore $\kap'$ is allowed to be positive (or zero), but a positive slope of the plastic strain profile
 is impossible to achieve without generating negative plastic strains inside the plastic zone near the boundary.
So this discussion leads to the conclusion that the condition $\kap'=0$ should be imposed at the boundary of the plastic
zone.
\end{itemize}

In the absence of body forces, we set $b=0$ and equation (\ref{eqgp106x})
implies that $EA(u'-\kap)$ is constant along the bar (independent of the
spatial coordinate $x$). Physically, this constant represents the axial force
transmitted by the bar and therefore will be denoted as $F$. 
Equation (\ref{eq:50}) is then rewritten as
\beql{eq:50mod}
HA(x)\kap(x)+(HA(x)l^2\kap'(x))'+A(x)\sig_0=F \mbox{ for } x\in\Ip
\eeq

\section{Bar With Piecewise Constant Stress Distribution}

\begin{figure}%
\centering
\begin{tabular}{cc}
(a) & (b) \\
\includegraphics[scale=0.28]{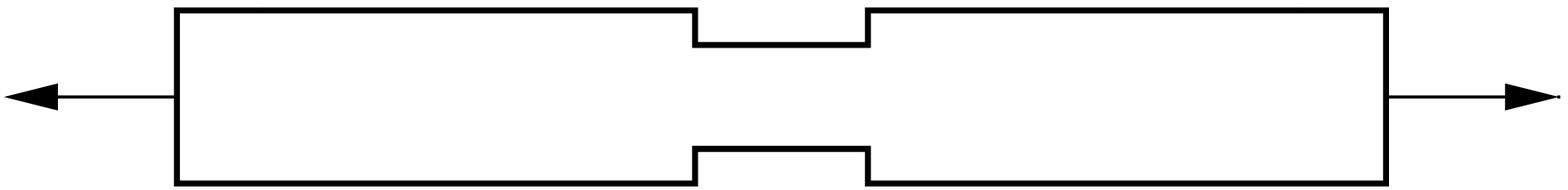}
&
\includegraphics[scale=0.28]{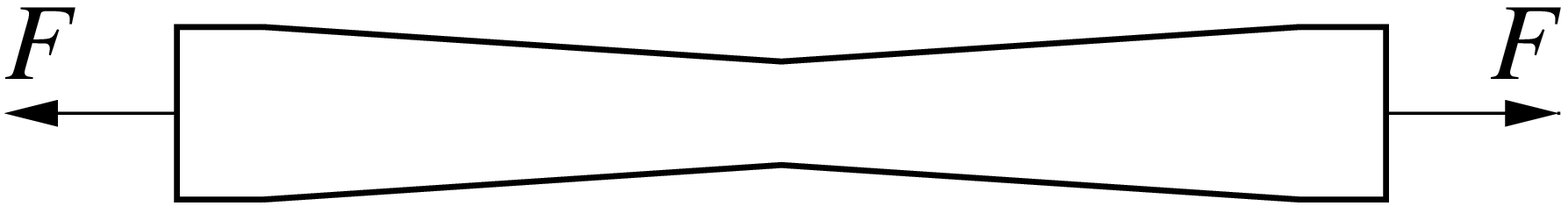}
\\[3mm]
& (c) 
\\
& \includegraphics[scale=0.28]{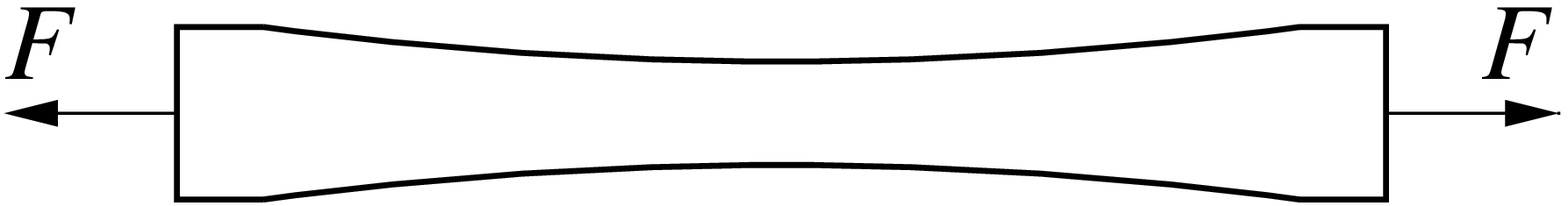} 
\end{tabular}
\caption{Tensile bars with (a) discontinuous distribution of stress, 
(b) continuous distribution of stress, (c) smooth distribution of stress}%
\label{fgp1}%
\end{figure}

Having presented the governing equations, we can proceed to localization
analysis of a tensile bar with variable cross section. As the first case,
consider a bar with piecewise constant sectional area (Fig.~\ref{fgp1}a). 
Suppose that
the bar contains a weak segment of length $2\lg$ and sectional area $A_c$, while
the remaining parts have a larger
sectional area $A_c/(1-\beta)$ where $\beta\in[0,1)$ is a dimensionless
parameter.
The origin of the spatial coordinate system can be placed into the center
of the weak segment. The solution is then expected to exhibit symmetry
with respect to the origin. 

Let us emphasize that the present analysis is strictly focused on
one-dimensional modeling. Therefore, the stress distribution across each
section is considered as uniform. Of course, for a real three-dimensional
body containing notches, the stress field would have a singularity at the notch
tip and the stress distribution across sections near that singularity would be
highly nonuniform. However, we use the case of variable cross section as
a paradigmatic example of the localization properties of a gradient
plasticity model in cases with non-smooth and sometimes even discontinuous
data.  
Therefore, the stress is expressed simply as the normal force divided
by the sectional area. 

\begin{figure}%
\centering
\includegraphics[scale=0.4]{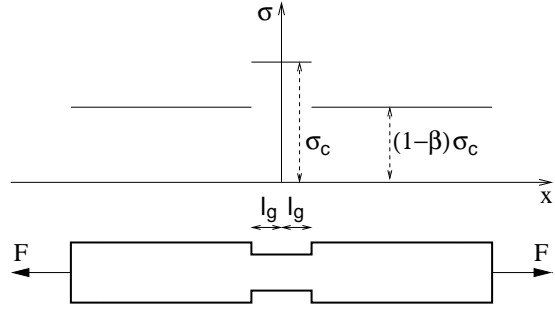}
\caption{Bar with a weak segment and the corresponding stress distribution}%
\label{fgp1x}%
\end{figure}

For the bar with a weak segment of length $2\lg$,
the  stress distribution is described by 
\beq
\sig(x) = \left\{ 
\begin{array}{ll}
F/A_c = \sigc & \mbox{ for } \vert x\vert < \lg \\
F/[A_c/(1-\beta)] = (1-\beta)\sigc & \mbox{ for } \vert x\vert > \lg
 \end{array}\right.
\label{eqgp10}
\eeq
and has discontinuities at sections $x=\pm\lg$; see Fig.~\ref{fgp1x}.
As long as the stress remains below the yield limit, the response
is purely elastic. The onset of yielding can be expected when the
yield limit is attained, which happens at the elastic limit force
$F_0=A_c\sigma_0$. For a softening plasticity model without any
regularization, plastic yielding would localize into one cross section,
the dissipation would vanish (because the plastic zone has zero volume)
and the response would be extremely brittle. Regularization by the 
additional gradient term leads to a finite size of the plastic zone. 

\subsection{Plastic Zone Contained in Weak Segment}\label{sec2.1}

Let us first assume that the plastic zone $\Ip$ is fully contained in 
the weak segment, i.e., $\Ip\subset(-\lg,\lg)$.
 The yield condition (\ref{eq:50mod}) with $A=A_c$ and $F=A_c\sigc$
can be rewritten as
\beql{eq:50z}
l^2\kap''(x)+\kap(x)=\frac{\sigc-\sig_0}{H} \mbox{ for } x\in\Ip
\eeq
This is a linear second-order differential equation with constant
coefficients and a constant right-hand side, and its general solution is
\beql{varfor11}
\kap(x) = \frac{\sigc-\sig_0}{H} + C_1 \cos\frac{x}{l} + C_2\sin\frac{x}{l}
\eeq
where $C_1$ and $C_2$ are arbitrary constants.
Let $\Lp$ denote the length of the plastic zone and suppose that the
plastic zone is centered at the origin, i.e., $\Ip=(-\Lp/2,\Lp/2)$.
If this was not the case, the origin could simply be shifted to the
center of the plastic zone.
As explained at the end of Section \ref{sec:varfor}, 
the plastic strain $\kap$ must remain continuous
and its derivative must vanish at the
elasto-plastic interface, $\partial\Ip$, which consists of two points,
$x=\pm\Lp/2$. Conditions $\kap(-\Lp/2)=0$,
$\kap(\Lp/2)=0$, $\kap'(-\Lp/2)=0$ and 
$\kap'(\Lp/2)=0$ lead to $C_2=0$, $C_1=(\sig_0-\sig_c)/(H\cos(\Lp/2l))$
and $\sin(\Lp/2l)=0$. The last condition means that the length of the
plastic zone, $\Lp$, must be an integer multiple of $2\pi l$. The shortest
positive length of plastic zone is obtained if $\Lp=2\pi l$. However,
such a solution is admissible only if the plastic zone is indeed fully
contained in the weak segment of length $2\lg$, i.e., if $\lg > \pi l$.
In that case, the solution is given by
\beql{varfor12}
\kap(x) = \left\{
\begin{array}{ll}
\displaystyle\frac{\sigc-\sig_0}{H} \left(1+\cos\frac{x}{l}\right) & \mbox{ for } x\in\Ip=(-\pi l,\pi l)\\
0 & \mbox{ for } x\notin\Ip
\end{array}\right.
\eeq 
This is the classical solution that describes localization in a bar
with perfectly uniform properties \cite{bormuhl92}. 
Analysis of the second variation of functional $\Pi$, presented
 in detail in Appendix~A, reveals that this solution corresponds to a local
minimum of $\Pi$.

If the weak segment is sufficiently
long with respect to the characteristic length of the material
(longer than $2\pi l$), the plastic zone will form inside that segment
and the solution will not be affected by stronger parts of the bar.
However, if the weak segment is shorter than the plastic zone in a perfectly
uniform bar, solution (\ref{varfor12}) is not admissible and the
derivation must be modified.

It is convenient and elegant to convert the problem to a dimensionless format.
For this purpose, we introduce the dimensionless spatial coordinate,
$\xi = x/l$, normalized plastic strain, $\kapt=-H\kappa/\sigma_0$, and
dimensionless stress $\phi=\sigc/\sig_0$. Note that $\phi$, defined as 
the ratio between the stress in the weak segment and the yield stress, 
is at the same time the ratio between the axial force, $F$, and its elastic
limit value, $F_0$, and thus will be referred to as the load parameter.
The distribution of plastic strain in a uniform bar, given by (\ref{varfor12}),
can be described in terms of the dimensionless quantities as
\beql{varfor13}
\kapt(\xi) = \left\{
\begin{array}{ll}
(1-\phi) \left(1+\cos\xi\right) & \mbox{ for } \xi\in(-\pi ,\pi)\\
0 & \mbox{ for } \xi\notin(-\pi,\pi)
\end{array}\right.
\eeq 
This solution is valid for a bar with a weak segment of length $2\lg$
provided that $2\lg>2\pi l$, i.e., $\lamg>\pi$ where
$\lamg=\lg/l$ is a dimensionless parameter describing the ratio
between the ``geometric'' characteristic length, $\lg$, and the material
characteristic length, $l$.
 
\subsection{Plastic Zone Extending to Strong Segments}

Let us proceed to the case when $2\lg< 2\pi l$, i.e., $\lamg<\pi$. 
Equation (\ref{eq:50z}) is now valid only for $\vert x\vert<\lg$.
In terms of the dimensionless quantities, we rewrite it as
\beql{eq:50y}
\kapt''(\xi)+\kapt(\xi)=1-\phi, \hskip 10mm \vert\xi\vert<\lamg
\eeq
For simplicity, the derivatives with respect to $\xi$ will still be denoted
by primes, despite possible confusion with derivatives with respect to $x$.
For the parts of the plastic zone surrounding the weak segment, 
a similar equation with a modified  right-hand side can be derived
from (\ref{eq:50mod}):
\beql{eq:50w}
\kapt''(\xi)+\kapt(\xi)=1-\phi+\beta\phi, \hskip 10mm \lamg<\vert\xi\vert<\lamp
\eeq
Here, $\lamp=\Lp/2l$ is a dimensionless parameter characterizing the ratio
between one half of the plastic zone length, $\Lp/2$, and the material
characteristic length, $l$. 
Since the solution is again expected to be symmetric with respect to the
origin, we construct it only in the weak segment
and in one of the stronger segments adjacent to it:
\beql{eqgp102}
\kapt(\xi)=\left\{\begin{array}{ll}
1-\phi +C_1\cos\xi+C_2\sin\xi & \mbox{ for } -\lamg\le \xi\le\lamg
\\[3mm]
1-\phi+\beta\phi +C_3\cos\xi+C_4\sin\xi & \mbox{ for } \lamg\le\xi\le\lamp
\end{array}\right.
\eeq
Integration constant $C_2$ must vanish due to symmetry. Constants $C_1$, $C_3$ and $C_4$ and the dimensionless plastic zone size $\lamp$
can be determined from four conditions: continuity of $\kapt$ and of $A\kapt'$ at $\xi=\lamg$ and at $\xi=\lamp$.
These conditions lead to the set of four equations
\bea
C_1\cos\lamg -C_3\cos\lamg-C_4\sin\lamg &=& \beta\phi \\
-C_1(1-\beta)\sin\lamg +C_3\sin\lamg-C_4\cos\lamg&=& 0 \\
C_3\cos\lamp+C_4\sin\lamp &=& (1-\beta)\phi - 1 \\
-C_3\sin\lamp+C_4\cos\lamp &=& 0
\eea
which are linear in terms of $C_1$, $C_3$ and $C_4$ but nonlinear in terms
of $\lamp$. Since the load parameter, $\phi$, enters the equations in
a linear fashion, it is of advantage to reformulate the problem and solve
for  the integration constants and $\phi$ in terms of $\lamp$. Another parameter
that affects the solution is $\lamg$, and so we can parameterize the solution
by $\lamg$ and $\lamp$ and write
\bea
\label{eqgp103}
C_1(\lamg,\lamp) &=& \beta\phi\,\frac{\sin(\lamp-\lamg)}{D(\lamg,\lamp)} \\
C_3(\lamg,\lamp) &=& (\beta-1)\beta\phi\,\frac{\sin\lamg\cos\lamp}{D(\lamg,\lamp)} \\
C_4(\lamg,\lamp) &=& (\beta-1)\beta\phi\,\frac{\sin\lamg\sin\lamp}{D(\lamg,\lamp)} 
\label{eqgp105}
\\
\label{eqgp106}
\phi(\lamg,\lamp) &=& \frac{1}{1-\beta}\frac{1}{1+\displaystyle\frac{\beta\sin\lamg}{D(\lamg,\lamp)}}
\eea
where 
\bea\nonumber
D(\lamg,\lamp)&=&(1-\beta\sin^2\lamg)\sin\lamp-\beta\sin\lamg\cos\lamg\cos\lamp=\\
&=& (1-\half\beta)\sin\lamp+\half\beta\sin(\lamp-2\lamg)
\eea

\begin{figure}%
\centering
\includegraphics[scale=0.6]{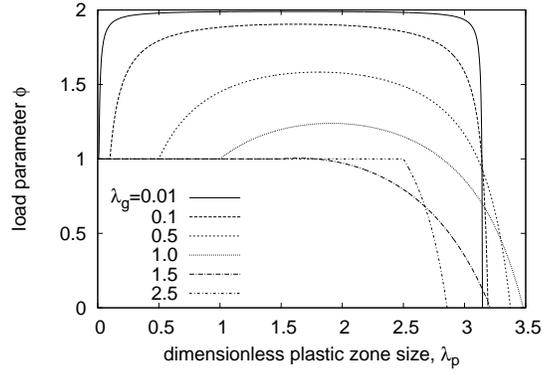}%
\caption{Relation between load parameter and plastic zone size for a piecewise constant stress distribution (bar with a weak segment)}%
\label{fig:aif2C-dl}%
\end{figure}

Recall that parameter $\phi$, defined as the ratio $\sigc/\sig_0$, can also
be interpreted as the ratio between the axial force transmitted by the bar,
$F=A_c\sigc$, and its limit elastic value, $F_0=A_c\sig_0$.
For a fixed $\lamg$, the dimensionless size of the plastic zone $\lamp$
can be varied as a parameter describing various stages of plastic zone 
evolution. The range in which the solution makes physical sense can
be determined from the conditions that the initial value of the load
parameter at the onset of yielding is $\phi=1$ and that $\phi=0$
corresponds to complete failure. 
The relation between the load parameter $\phi$ and the dimensionless plastic zone size $\lamp$ is graphically
illustrated in Fig.~\ref{fig:aif2C-dl} for $\beta=0.5$ and for several values of parameter $\lamg$
ranging from 0.01 to 2.5. The interesting range of $\lamg$ is between zero and $\pi$
because for $\lamg>\pi$ the weak zone is long enough to allow formation of the complete plastic zone in its interior,
and the solution is the same as for a bar with a perfectly uniform section $A_c$.

When the axial force
reaches the limit elastic value (i.e., when $\phi=1$),
the entire weak segment of length $2\lg$ starts yielding and the initial value of the dimensionless plastic
zone size is $\lamp=\lamg$. Subsequently, the plastic zone expands continuously, and this initially happens at an
increasing load level, provided that $\lamg<\pi/2$, i.e., that the length of the weak segment $2\lg$ is smaller
than $\pi l$. The maximum force 
\beq
F_{max} = \frac{F_0}{1-\beta}\frac{1}{1+\displaystyle\frac{\beta\sin\lamg}{\sqrt{1-\beta(2-\beta)\sin^2\lamg}}}
\eeq
is attained when the dimensionless size of the plastic zone is
\beq
\lambda_{p,peak}= \pi-\arctan\frac{1-\beta\sin^2\lamg}{\beta\sin\lamg\cos\lamg}
\eeq
After that, the axial force decreases to zero as the size of the plastic zone approaches the limit
\beq
\lambda_{p,max} = \lambda_{p,peak} + \half\pi 
\eeq
If the size of the weak segment exceeds $\pi l$, the global response is softening right from the onset
of plastic yielding.

\begin{figure}%
\centering
\begin{tabular}{cc}
(a) & (b)
\\
\hskip -20mm
\includegraphics[scale=0.6]{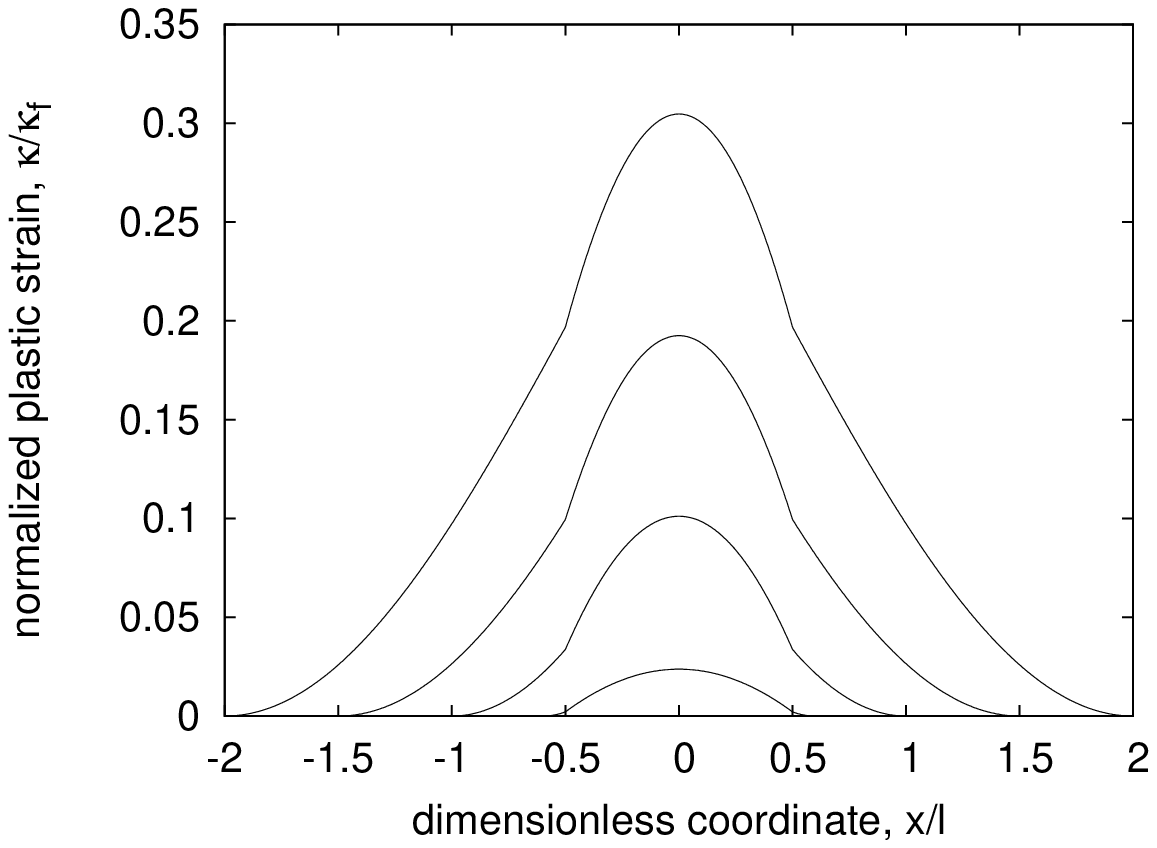}%
&
\includegraphics[scale=0.6]{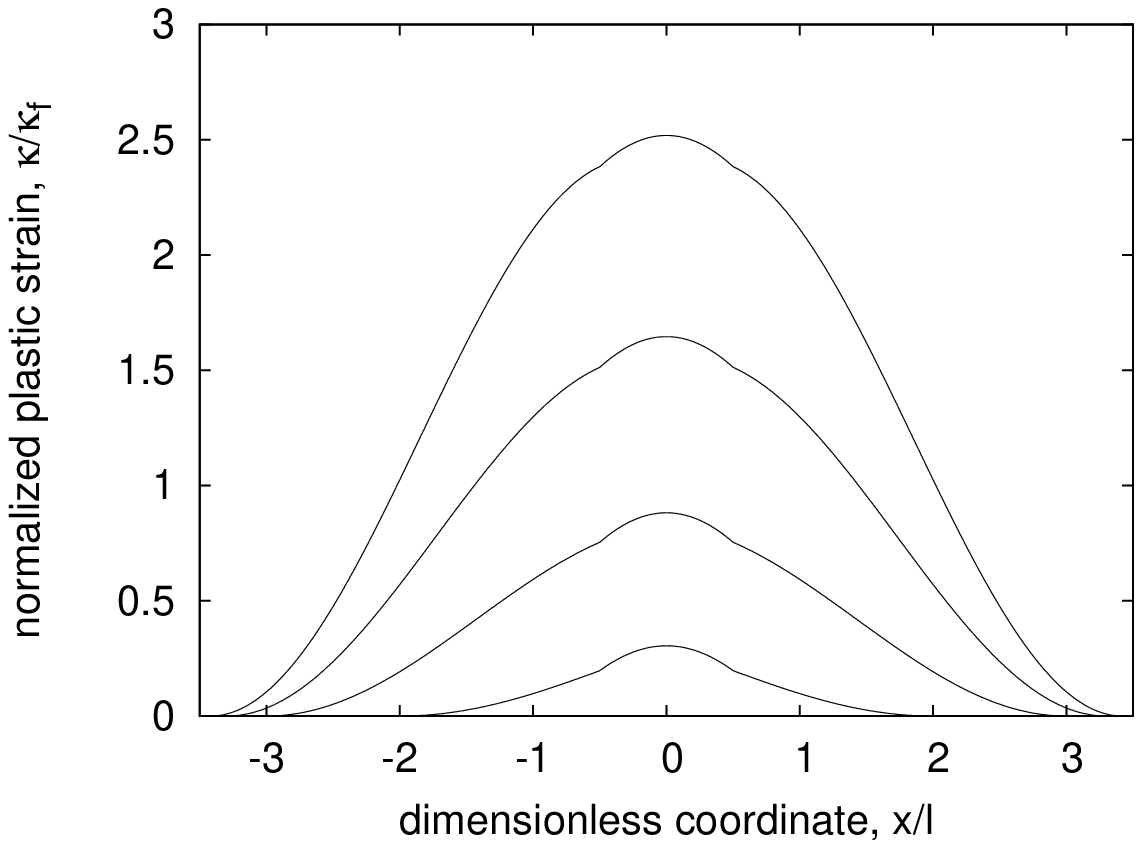}%
\end{tabular}
\caption{(a) Early stage, (b) late stage of evolution of plastic strain profile for a piecewise constant stress distribution}%
\label{fig:aif2C-kap}%
\end{figure}

Substituting from (\ref{eqgp103})--(\ref{eqgp106}) into (\ref{eqgp102}), we obtain the distribution of plastic
strain parameterized by $\lamp$, and integrating over the plastic zone we get the plastic elongation.
An example of the evolution of plastic strain profile, calculated for $\beta=0.5$ and $\lamg=0.5$,
is shown in Fig.~\ref{fig:aif2C-kap}. The plastic zone expands and, at each section, the plastic strain grows monotonically.

It is also of interest to construct the load-displacement diagram
for the entire bar. 
The elastic elongation is proportional to the axial force and the
proportionality factor (bar compliance) depends on the bar length.
Therefore, it is convenient to consider the contribution of plastic strain
to the elongation separately, since the bar length does not affect it
(provided that the bar is  sufficiently long such that the 
full plastic zone can develop). 
The plastic part of bar elongation,
\beq
u_p = \int_{-\Lp/2}^{\Lp/2} \kappa(x)\;\dx
\eeq
can be computed by
integrating the plastic strain along the plastic zone.
In the context of dimensionless description, it is natural to deal with the 
dimensionless plastic elongation
\beq
 \int_{-\lamp}^{\lamp} \kapt(x)\;\dxi=
\int_{-\Lp/2}^{\Lp/2} -\frac{H\kappa(x)}{\sigma_0}\frac{\dx}{l} =-\frac{Hu_p}{\sigma_0l} =\frac{u_p}{\kappa_f l}
\eeq
where $\kappa_f=-\sigma_0/H$ is a material parameter that corresponds 
to the plastic strain at complete failure if the gradient terms are ignored. 

\begin{figure}%
\centering
\begin{tabular}{cc}
(a) & (b)
\\
\hskip -20mm
\includegraphics[scale=0.6]{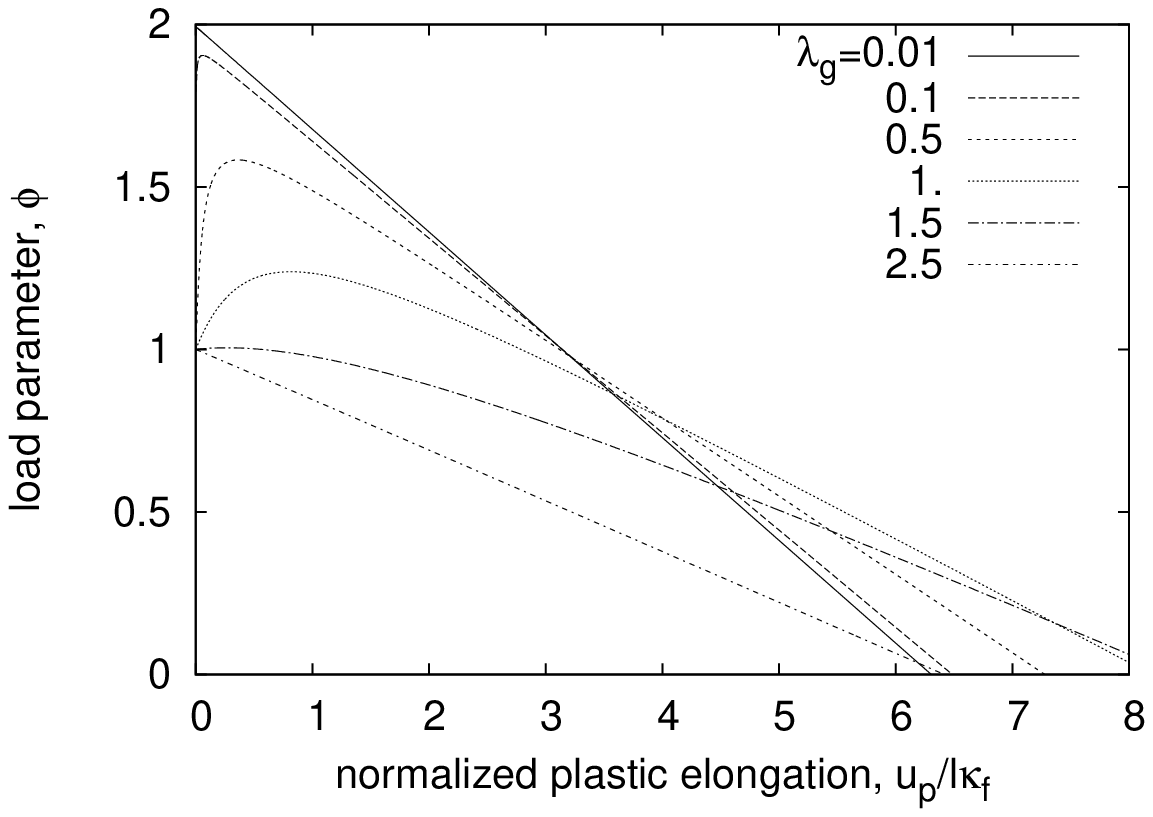}%
&
\includegraphics[scale=0.6]{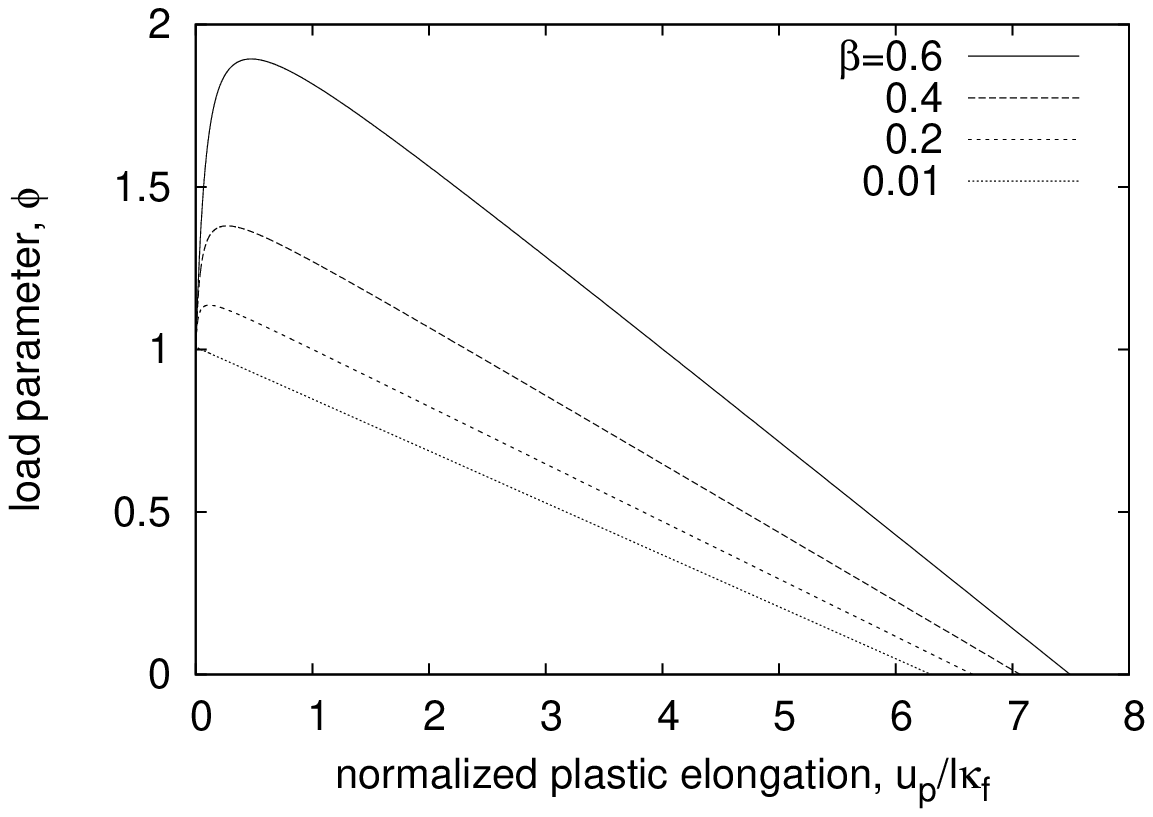}%
\end{tabular}
\caption{Plastic part of load-displacement diagram for a piecewise constant stress distribution for (a) different values of relative imperfection size $\lamg=\lg/l$, (b) different values of imperfection ``magnitude'' $\beta$}%
\label{fig:aif2C-sd}%
\end{figure}

The plastic part of the load-displacement diagram, obtained by plotting the 
dimensionless load parameter $\phi$ against the dimensionless plastic elongation
$u_p/\kappa_f l$, 
is shown in Fig.~\ref{fig:aif2C-sd}a for fixed $\beta=0.5$ and different
dimensionless sizes of the weak segment, $\lamg$, and in Fig.~\ref{fig:aif2C-sd}b for fixed $\lamg=0.5$ and different values of $\beta$.
These graphs explain the influence of both parameters on the shape of the load-displacement diagram.
The weak segment can be considered as an imperfection in a uniform bar, parameter $\lamg$ refers to the length
of that imperfection and $\beta$ to its ``magnitude'' (in the sense that larger $\beta$ means a more dramatic reduction
of the sectional area). 
For short imperfections, the initial structural hardening is very steep and the maximum axial force is close to the
value calculated for a perfect bar. For somewhat longer imperfections, the structural hardening is more gradual and the maximum
force is between the values that would correspond to the weaker and stronger sections, and for imperfection lengths
exceeding $\pi l$ (which is exactly one half of the plastic zone that would form in a perfectly uniform bar)
the maximum load is dictated exclusively by the weak section and the load-displacement diagram is softening
from the onset of yielding.

\section{Bar With Piecewise Linear Stress Distribution}

\subsection{General Solution}

Now consider a bar with variable sectional area described by a function
that is continuous but not continuously differentiable at the weakest section;
see Fig.~\ref{fgp1}b. To allow for analytical solutions, we define
the specific distribution of sectional area such that the corresponding stress
distribution becomes piecewise linear. This is achieved by setting
\begin{equation}
A(x)=\frac{A_cl_g}{l_g-|x|}
\label{eqgp2}
\end{equation}
where $\lg$ is a parameter that sets the geometric length scale of the problem.
Substituting (\ref{eqgp2}) into (\ref{eq:50mod}) and dividing by $HA(x)$, 
we obtain
\beql{eqgp2x}
  l^2\kappa''(x)+\frac{l^2\sgn{x}}{l_g-|x|}\kappa'(x) +\kappa(x) =\frac{\sigma_c(l_g-|x|)}{Hl_g}-\frac{\sigma_0}{H} \hskip 10mm \mbox{ for } x\in\Ip-\{0\}
\eeq
Equation (\ref{eqgp2x}) must be satisfied at all points inside the plastic zone,
with the exception of point $x=0$ at which the sectional area is not
differentiable. At that point, conditions of continuity of $\kappa$ and $A\kap'$
have to be imposed. Since $A$ is continuous, the latter condition actually
means continuity of $\kap'$. After conversion to dimensionless form,
the governing equation reads
\begin{equation}
\kapt''(\xi) +\frac{\sgn{\xi}}{\lambda_g-|\xi|}\kapt'(\xi)+\kapt(\xi) =1-\phi\,\left(1-\frac{|\xi|}{\lamg}\right)
\label{eqgp3}
\end{equation}
This is a second-order differential equation, but in contrast to (\ref{eq:50y})
or (\ref{eq:50w}),
it contains on the left-hand side an additional term with the first-order 
derivative which has a non-constant coefficient, and the right-hand side is
not constant. Still, an analytical solution in terms of special functions
can be constructed. The derivation is presented in detail in Appendix~B.
The resulting general solution has the form
\bea\nonumber
\kapt(\xi)&=&(\lamg-|\xi|)C_1^{\pm}\mbox{J}_1(\lamg-|\xi|)+(\lamg-|\xi|)C_2^{\pm}\mbox{Y}_1(\lamg-|\xi|)+
\\
&&+1-\frac{\phi\pi(\lamg-|\xi|)}{2\lambda_g}\mbox{\bf H}_1(\lamg-|\xi|)\quad\quad\quad \mbox{for}\:\xi\in\Ip
\label{eqgp20}
\eea
where $C_1^{\pm}$ and $C_2^{\pm}$ are integration constants, $\mbox{J}_1$ and $\mbox{Y}_1$ are
the Bessel functions of the first and second kind, resp., 
and $\mbox{\bf H}_1$ is the Struve function \cite{uvodbessel}. 

\subsection{Particular Solution}

For simplicity, we have written (\ref{eqgp20}) in a compact form, 
but the integration constants have different values in the ``positive part''
of the plastic zone, where $\xi>0$, and in the ``negative part'' of the
plastic zone, where $\xi<0$. Therefore, we deal with four integration
constants, $C_1^+$, $C_2^+$, $C_1^-$ and $C_2^-$, and with two additional
unknowns that determine the position of the right and left boundary of the
plastic zone. This makes  a total of six unknowns, which can be determined 
from appropriate jump and boundary conditions. According to the foregoing analysis,
the solution must remain continuously differentiable at the origin and
at the two boundary points, which leads to six equations for the six
unknowns. 

Due to symmetry, the problem can be simplified and it is
sufficient to restrict attention to the positive part of the plastic
zone, $\Ip^+=(0,\lamp)$. 
The three relevant unknowns, $C_1^+$, $C_2^+$ and $\lamp$, can be
determined from three conditions,
\bea\label{cond1}
\kapt'(0) &=& 0 \\
\kapt(\lamp) &=& 0\\
\kapt'(\lamp) &=& 0\label{cond3}
\eea
Substituting the general solution (\ref{eqgp20}) into (\ref{cond1})--(\ref{cond3}),
we obtain a set of three equations,
\bea
\hskip -20mm
-2\lambda_g\left[C_1^+\mbox{J}_0(\lambda_g)+C_2^+\mbox{Y}_0(\lambda_g)\right]+\phi\pi\mbox{\bf H}_0(\lambda_g)&=&0
\\
\hskip -20mm
C_1^+\mbox{J}_1(\lambda_g-\lambda_p)+C_2^+\mbox{Y}_1(\lambda_g-\lambda_p)-\frac{\phi\pi}{2\lambda_g}\mbox{\bf H}_1(\lambda_g-\lambda_p)+\frac{1}{\lambda_g-\lambda_p}&=&0
\\
\hskip -20mm
-2\lambda_g\left[C_1^+\mbox{J}_0(\lambda_g-\lambda_p)+C_2^+\mbox{Y}_0(\lambda_g-\lambda_p)\right]+\phi\pi\mbox{\bf H}_0(\lambda_g-\lambda_p)&=&0
\eea
which are linear in terms of $C_1^+$ and $C_2^+$ but nonlinear in terms
of $\lamp$. Again, it is of advantage to reformulate the problem and solve
for  $C_1^+$, $C_2^+$ and $\phi$ in terms of $\lamp$ and $\lamg$:
\bea
C_1^+(\lambda_g,\lambda_p)&=&\frac{\pi\left[-\mbox{Y}_0(\lambda_g-\lambda_p)\mbox{\bf H}_0(\lambda_g)+\mbox{Y}_0(\lambda_g)\mbox{\bf H}_0(\lambda_g-\lambda_p)\right]}{D^+(\lambda_g,\lambda_p)}
\\
C_2^+(\lambda_g,\lambda_p)&=&\frac{\pi\left[\mbox{J}_0(\lambda_g-\lambda_p)\mbox{\bf H}_0(\lambda_g)-\mbox{J}_0(\lambda_g)\mbox{\bf H}_0(\lambda_g-\lambda_p)\right]}{D^+(\lambda_g,\lambda_p)}
\\
\phi(\lambda_g,\lambda_p)&=&\frac{2\lambda_g\left[\mbox{J}_0(\lambda_g-\lambda_p)\mbox{Y}_0(\lambda_g)-\mbox{J}_0(\lambda_g)\mbox{Y}_0(\lambda_g-\lambda_p)\right]}{D^+(\lambda_g,\lambda_p)}
\eea
where
\bea
D^+(\lambda_g,\lambda_p)&=&2\mbox{\bf H}_0(\lambda_g)+\pi(\lambda_g-\lambda_p)\left\{\left[-\mbox{J}_1(\lambda_g-\lambda_p)\mbox{Y}_0(\lambda_g)+\right.\right.
\\ \nonumber
&&\left.+\mbox{J}_0(\lambda_g)\mbox{Y}_1(\lambda_g-\lambda_p)\right]\mbox{\bf H}_0(\lambda_g-\lambda_p)+
\\ \nonumber
&&+
\left.\left[\mbox{J}_0(\lambda_g-\lambda_p)\mbox{Y}_0(\lambda_g)-\mbox{J}_0(\lambda_g)\mbox{Y}_0(\lambda_g-\lambda_p)\right]\mbox{\bf H}_1(\lambda_g-\lambda_p)\right\}
\eea
From symmetry with respect to the origin, we obtain integration constants
corresponding to the negative part of the plastic zone as $C_1^-=C_1^+$
and $C_2^-=C_2^+$. The assumption of symmetry may seem to be restrictive,
but it can be verified by (tedious) analysis of the complete problem
with six equations and six unknowns that no nonsymmetric admissible
solution exists.

\subsection{Results and Discussion}

\begin{figure}[h]
\centering
\includegraphics[scale=0.6]{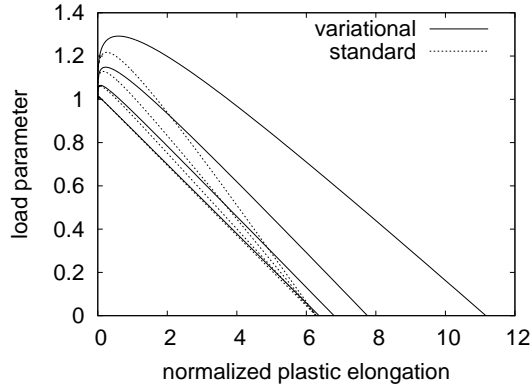}
\caption{Plastic part of normalized load-displacement diagram for piecewise linear stress distribution}
\label{fgp2}
\end{figure}

Based on the solution constructed in the previous subsection,
the evolution of the plastic zone can be analyzed, and the
corresponding load-displacement diagram can be constructed. 
The plastic part of the load-displacement diagram is shown
in Fig.~\ref{fgp2}, where the dimensionless load parameter $\phi$
is plotted against the dimensionless plastic elongation, $u_p/l\kappa_f$.
Recall that $\phi=\sigc/\sig_0=F/F_0$ is the ratio between the axial
force $F$ and its value at the onset of yielding, $F_0=A_c\sigma_0$.
The plastic elongation, $u_p$, is normalized by a reference value $l\kappa_f$,
which would correspond to the plastic elongation of a bar of length $l$
at complete failure if the solution remained uniform.

During the elastic stage of loading, the axial force increases from 0
to $F_0$ and no plastic strain evolves. Thus the initial part of the
diagram in Fig.~\ref{fgp2}, up to $\phi=1$, is vertical. For a bar with
a uniform section, the continuation of that diagram would be a straight line
that corresponds to linear softening, and the plastic elongation
at complete failure ($\phi=0$) would be $2\pi l\kappa_f$. This is in fact
a limit case of the present solution with $\lamg\rightarrow\infty$.
For finite $\lamg$, i.e., for a bar with a variable section, the
load parameter first increases and only later decreases. Complete
failure is attained at larger elongations than for the uniform bar.
This is represented in Fig.~\ref{fgp2} by the solid curves, which correspond
to different values of parameter $\lamg=3.2$, 5, 10 and 50 (from top 
to bottom). Lower values of $\lamg$ correspond to stronger variation
of the secional area and lead to higher peak loads and higher elongations
at failure. For comparison, the solutions constructed in \cite{jzv10}
using a ``standard'' formulation, with the second term in (\ref{eq:50mod})
replaced by $HAl^2\kappa''$, are plotted by the dotted curves. 
The present, variationally based formulation leads to qualitatively similar
solutions, but with higher peak loads    and higher elongations at complete
failure. Note that, for the standard formulation, the elongation
at failure is always the same as for a uniform bar, independently of
the value of parameter $\lamg$.

\begin{figure}[h]
\centering
\includegraphics[scale=0.6]{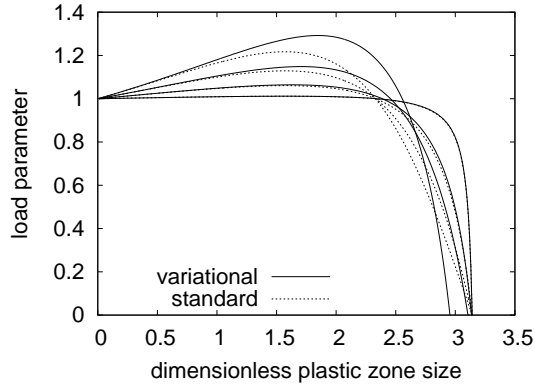}
\caption{Evolution of plastic zone size for piecewise linear stress distribution}
\label{fgp3}
\end{figure}

The evolution of the plastic zone size is illustrated in Fig.~\ref{fgp3}
by plotting the load parameter $\phi$ against the dimensionless plastic
zone size $\lamp = \Lp/2l$, again for $\lamg=3.2$, 5, 10 and 50.
 For a uniform bar ($\lamg\rightarrow\infty$), the plastic zone would
attain its full size $\Lp=2\pi l$ (corresponding to $\lamp=\pi$)
immediately at the onset of yielding. In contrast to that, for the
bar with a variable section, the plastic zone evolves continuously from 
the weakest  section up to its full size, attained at complete failure.
The fact that the plastic zone expands monotonically is important,
because it justifies our tacit assumption that the material outside
the current plastic zone has not experienced any plastic straining before.
The full size of the plastic zone  turns out to be somewhat smaller
than for a uniform bar, which is different from the standard solution
constructed in \cite{jzv10},
plotted for comparison by dotted curves.

\begin{figure}
\centering
\includegraphics[scale=0.6]{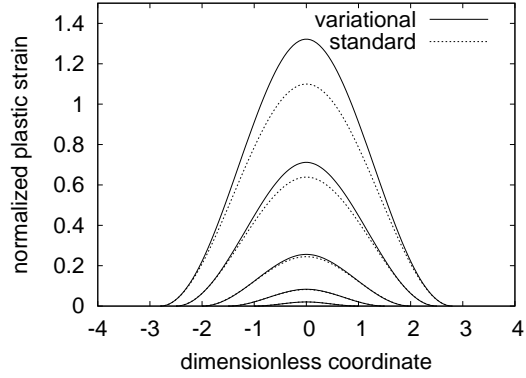}
\caption{Evolution of plastic strain profile for piecewise linear stress distribution}
\label{fgp4}
\end{figure}

The evolution of the plastic strain profile is plotted in Fig.~\ref{fgp4}
for $\lambda_g=5$. Again, solid curves represent the variational formulation
and dotted curves the standard formulation from \cite{jzv10}.
At early stages of plastic strain evolution, both formulations give
almost the same result, but later the differences grow. 



\section{Bar With Smooth Stress Distribution}

As the most regular case, we consider a smooth distribution
of sectional area, with continuous derivatives of an arbitrary order.
The origin of the spatial coordinate is again placed at the weakest
section, where plastic yielding is expected to start.
In \cite{jzv10}, the function describing the sectional area was
selected such that the resulting stress distribution was quadratic.
For the present purpose, it turns out to be more convenient to
consider another special case in which the stress distribution is given by
a Gaussian function,
\beq
\sigma(x) = \sigc\,\eee^{-{x^2}/{2l_g^2}}
\eeq
The sectional area is thus
\beq
A(x) = A_c\, \eee^{{x^2}/{2l_g^2}}
\eeq
 Since, for this case, 
the solution based on the standard formulation
has not been published yet, let us present it before turning attention
to the variational formulation.

\subsection{Solution Based on Standard Formulation}\label{sec4.1}

According to the standard formulation used in \cite{jzv10}, 
the yield condition is considered in the form
\beql{eq:50q}
HAl^2\kap''(x)+HA\kap(x)+A\sig_0=F \mbox{ for } x\in\Ip
\eeq
which differs from (\ref{eq:50mod}) by the second term.
The corresponding differential equation for the plastic strain
can be converted into the dimensionless form
\begin{equation}
\kapt''(\xi)+\kapt(\xi)=1-\phi\, \eee^{-{\xi^2}/{2\lambda_g^2}}
\label{eqgp125}
\end{equation}
The general solution of the corresponding homogeneous equation
is a linear combination of $\cos\xi$ and $\sin\xi$, and the particular
solution for the given right-hand side can be constructed by variation
of constants. For the particular solution represented by
\beq\label{eqgp125y}
\tilde{\kappa}_{\rm n}(\xi)=\tilde{C}_1(\xi)\cos{\xi}+\tilde{C}_2(\xi)\sin{\xi}
\eeq
we obtain a set of two equations
\bea
\tilde{C}_1'(\xi)\cos{\xi}+\tilde{C}_2'(\xi)\sin{\xi}&=&0
\\
-\tilde{C}_1'(\xi)\sin{\xi}+\tilde{C}_2'(\xi)\cos{\xi}&=&1-\phi \,\eee^{-{\xi^2}/{2\lambda_g^2}}
\eea
from which
\bea
\label{eqc1}
\tilde{C}_1(\xi) &=& \cos\xi+\phi\int \eee^{-{\xi^2}/{2\lambda_g^2}}\sin\xi\, 
\dxi
\\
\label{eqc2}
\tilde{C}_2(\xi) &=& \sin\xi-\phi\int \eee^{-{\xi^2}/{2\lambda_g^2}}\cos\xi\, 
\dxi
\eea
The integrals can be conveniently expressed by switching to complex
functions. As shown in Appendix~C, the resulting general solution
of equation (\ref{eqgp125}) can be expressed as 
\beq\label{eqdaws}
\kappa_{\rm n}(\xi)=C_1\cos{\xi}+C_2\sin{\xi}+1-\phi \lamg\eee^{-\xi^2/2\lamg^2}\frac{\sqrt{2}}{2}
\left[F\left(\frac{\lamg^2-i\xi}{\sqrt{2}\lamg}\right)+F\left(\frac{\lamg^2+i\xi}{\sqrt{2}\lamg}\right)\right]
\eeq
where $C_1$ and $C_2$ are arbitrary constants, $i$ denotes the imaginary unit,
 and
\beq\label{defdaw}
\mbox{F}(x)=\eee^{-x^2}\int\limits_0^x \eee^{t^2}\mbox{d}t
\eeq
is the so-called Dawson function; see e.g.\ \cite{olver}.

In general, one should impose two boundary conditions (of vanishing value
and vanishing derivative) at each boundary point and solve for two
integration constants $C_1$ and $C_2$ and two unknown coordinates of the
boundary points. Due to symmetry, the problem can be simplified.
Integration constant $C_2$ must vanish, and the plastic zone 
$\Ip=(-\lamp,\lamp)$ is characterized by one unknown parameter, $\lamp$.
Boundary conditions $\kapt(\lamp)=0$ and $\kapt'(\lamp)=0$ lead to
a set of two equations,
\bea
\hskip -30mm
-C_1\cos{\lambda_p}
+\phi \frac{\lamg}{\sqrt{2}}\eee^{-\lamp^2/2\lamg^2}
\left[F\left(\frac{\lamg^2-i\lamp}{\sqrt{2}\lamg}\right)+F\left(\frac{\lamg^2+i\lamp}{\sqrt{2}\lamg}\right)\right]
&=&1
\label{eqgp128}
\\
\hskip -30mm
C_1\sin{\lambda_p}+\phi\frac{i\lambda_g}{\sqrt{2}}\,\eee^{-\frac{\lambda_p^2}{2\lambda_g^2}}\left[\mbox{F}\left(\frac{\lambda_g^2-i\lambda_p}{\lambda_g\sqrt{2}}\right)-\mbox{F}\left(\frac{\lambda_g^2+i\lambda_p}{\lambda_g\sqrt{2}}\right)\right]&=&0
\label{eqgp128xx}
\eea
which can be written as
\bea
-c_p C_1 +EF_R\phi &=& 1 \\
s_p C_1 + EF_I\phi &=& 0
\eea
where $c_p=\cos\lamp$, $s_p=\sin\lamp$, $E=\sqrt{2}\lamg\eee^{-\lamp^2/2\lamg^2}$
and
\bea
F_R &=& \frac{1}{2}\left[\mbox{F}\left(\frac{\lambda_g^2+i\lambda_p}{\lambda_g\sqrt{2}}\right)+\mbox{F}\left(\frac{\lambda_g^2-i\lambda_p}{\lambda_g\sqrt{2}}\right)\right] \\
F_I &=& -\frac{i}{2}\left[\mbox{F}\left(\frac{\lambda_g^2+i\lambda_p}{\lambda_g\sqrt{2}}\right)-\mbox{F}\left(\frac{\lambda_g^2-i\lambda_p}{\lambda_g\sqrt{2}}\right) \right]
\eea
are the real and imaginary parts of $\mbox{F}\left(\frac{\lambda_g^2+i\lambda_p}{\lambda_g\sqrt{2}}\right)$.
Equation (\ref{eqgp128xx}) follows from the condition $\kapt'(\lamp)=0$ 
with the derivative of $\mbox{F}$ expressed as $\mbox{F}'(x)=1-2x\mbox{F}(x)$,
which easily follows from the definition of Dawson function (\ref{defdaw}).

Solving for $C_1$ and $\phi$, we obtain
\bea
C_1(\lamp,\lamg)&=&-\frac{F_I}{F_Rs_p+F_Ic_p}=
\frac{\mbox{F}\left(\frac{\lambda_g^2+i\lambda_p}{\lambda_g\sqrt{2}}\right)-\mbox{F}\left(\frac{\lambda_g^2-i\lambda_p}{\lambda_g\sqrt{2}}\right)}{\eee^{-i\lambda_p}\mbox{F}\left(\frac{\lambda_g^2-i\lambda_p}{\lambda_g\sqrt{2}}\right)-\eee^{i\lambda_p}\mbox{F}\left(\frac{\lambda_g^2+i\lambda_p}{\lambda_g\sqrt{2}}\right)}
\label{eqgp129}
\\
\nonumber
\phi(\lamp,\lamg)&=&\frac{s_p}{E(F_Rs_p+F_Ic_p)}=
\frac{\sqrt{2}}{\lamg}
\frac{i\sin\lamp \eee^{\lamp^2/2\lamg^2}}
{\eee^{i\lamp}\mbox{F}\left(\frac{\lambda_g^2+i\lambda_p}{\lambda_g\sqrt{2}}\right)-\eee^{-i\lamp}\mbox{F}\left(\frac{\lambda_g^2-i\lambda_p}{\lambda_g\sqrt{2}}\right)}
\\
\label{eqgp129x}
\eea

As usual, the solution is parameterized by the dimensionless size
of the plastic zone, $\lamp$. For a given value of $\lamp$, the corresponding
load parameter $\phi$ is given by (\ref{eqgp129x}) and the distribution
of plastic strain is obtained by substituting $C_1$ given by (\ref{eqgp129})
and $C_2=0$ into (\ref{eqdaws}).

\subsection{Solution Based on Variational Formulation}\label{sec4.2}

For the variational formulation, the yield condition is used in the
form (\ref{eq:50mod}) instead of (\ref{eq:50q}), and
equation (\ref{eqgp125}) is replaced by
\begin{equation}
\eee^{-{\xi^2}/{2\lambda_g^2}}\left(\eee^{{\xi^2}/{2\lambda_g^2}}\kappa_n(\xi)'\right)'+\kappa_n(\xi)=1-\phi \,\eee^{-{\xi^2}/{2\lambda_g^2}}
\label{eqgp101}
\end{equation}
As shown in Appendix~D,
the general solution of the corresponding homogeneous equation is a linear
combination of functions
\ignore
{
\bea
\kappa_{n,1}(\xi)&=&\eee^{-{\xi^2}/{2\lambda_g^2}}\mbox{H}_{(\lambda_g^2-1)}\left(\frac{\xi}{\lambda_g\sqrt{2}}\right)
\label{eqgp114}
\\
\kappa_{n,2}(\xi)&=&\eee^{-{\xi^2}/{2\lambda_g^2}}{_1}\;\mbox{F}_1\left(\frac{1}{2}(1-\lambda_g^2),\frac{1}{2},\frac{\xi^2}{2\lambda_g^2}\right)
\label{eqgp115}
\eea
}
\bea
\kappa_{n,1}(\xi)&=&\eee^{-\frac{\xi^2}{2\lambda_g^2}}\; {_1}\hskip -0.5mm \mbox{F}_1\left(\frac{1-\lambda_g^2}{2};\frac{1}{2};\frac{\xi^2}{2\lambda_g^2}\right)
\label{eqgp114}
\\
\kappa_{n,2}(\xi)&=&\eee^{-\frac{\xi^2}{2\lambda_g^2}}\frac{\xi}{\sqrt{2}\lambda_g} {_1}\hskip -0.5mm \mbox{F}_1\left(\frac{2-\lambda_g^2}{2};\frac{3}{2};\frac{\xi^2}{2\lambda_g^2}\right)
\label{eqgp115}
\eea
where ${_1}\hskip -0.5mm \mbox{F}_1$ denotes the so-called confluent hypergeometric function of the first kind \cite{Sneddon}, 
defined by formulae (\ref{eq11})--(\ref{eq12}) in Appendix~D. 
A particular solution of the non-homogeneous equation (\ref{eqgp101}) 
could be constructed
by variation of constants. Following the same procedure as in Section 
\ref{sec4.1}, we obtain a particular solution
\beq
\tilde{\kappa}_{\rm n}(\xi) = \tilde{C}_1(\xi)\kappa_{n,1}(\xi)+\tilde{C}_2(\xi)\kappa_{n,2}(\xi)
\eeq
with
\bea
\hskip -10mm
\tilde{C}_1(\xi) &=& \int\frac{\kappa_{n,2}(\xi)}{\kappa'_{n,1}(\xi)\kappa_{n,2}(\xi)-\kappa_{n,1}(\xi)\kappa'_{n,2}(\xi)}\left(1-\phi\, \eee^{-\frac{\xi^2}{2\lambda_g^2}}\right)\dxi
\\
\hskip -10mm
\tilde{C}_2(\xi) &=& -\int\frac{\kappa_{n,1}(\xi)}{\kappa'_{n,1}(\xi)\kappa_{n,2}(\xi)-\kappa_{n,1}(\xi)\kappa'_{n,2}(\xi)}\left(1-\phi \,\eee^{-\frac{\xi^2}{2\lambda_g^2}}\right)\dxi
\eea
Unfortunately, these integrals cannot be evaluated analytically.

An alternative approach can be based on the Green function
of the differential operator on the left-hand side of (\ref{eqgp101}).
Formally, the Green function represents the solution of the differential
equation with the right-hand side replaced by Dirac distribution centered
at a point $\eta$. At points $\xi$ different from $\eta$, the right-hand
side is zero and the solution is a linear combination of functions
(\ref{eqgp114})--(\ref{eqgp115}). 
Due to the singularity at $\xi=\eta$, different coefficients
of linear combination must be used for $\xi\le\eta$ and for $\xi\ge\eta$.
Moreover, these coefficients depend on the specific value of $\eta$.
Therefore, we can write the Green function as
\beq\label{green0}
G(\xi,\eta) = \left\{ \begin{array}{ll}
A_1(\eta)\kappa_{n,1}(\xi)+A_2(\eta)\kappa_{n,2}(\xi) & \mbox{ for } -\lamp\le\xi\le\eta
\\
B_1(\eta)\kappa_{n,1}(\xi)+B_2(\eta)\kappa_{n,2}(\xi) & \mbox{ for } \eta\le\xi\le\lamp
\end{array}\right.
\eeq
and impose at $\xi=\eta$ the continuity condition for the value and the unit
jump condition for the first derivative. This leads to two equations,
\bea\label{green1}
\hskip -10mm
A_1(\eta)\kappa_{n,1}(\eta)+A_2(\eta)\kappa_{n,2}(\eta) &=& B_1(\eta)\kappa_{n,1}(\eta)+B_2(\eta)\kappa_{n,2}(\eta)
\\
\hskip -10mm
A_1(\eta)\kappa'_{n,1}(\eta)+A_2(\eta)\kappa'_{n,2}(\eta) &=& B_1(\eta)\kappa'_{n,1}(\eta)+B_2(\eta)\kappa'_{n,2}(\eta)-1
\label{green2}
\eea
In usual problems solved on a fixed interval, the Green function should 
also satisfy two boundary conditions (one at each boundary point).
However, in our case the exact position of the boundary points is not
specified and the number of conditions to be satisfied at each boundary
point is two (vanishing value and vanishing first derivative). 
Making use of symmetry, we can restrict attention to non-negative
values of $\xi$ and $\eta$ and impose only one condition of vanishing
derivative at $\xi=0$, while at $\xi=\lamp$ we still need to satisfy two
conditions. One of them can be incorporated into the Green function, and the
other will be imposed aposteriori on the resulting solution and will
provide a link between the load parameter $\phi$ and the dimensionless
size of the plastic zone, $\lamg$.

Based on the foregoing discussion, we constrain the Green function by
conditions of vanishing derivative at $\xi=0$ and vanishing value
at $\xi=\lamp$, which gives two additional equations,
\bea\label{green4a}
A_1(\eta)\kappa'_{n,1}(0)+A_2(\eta)\kappa'_{n,2}(0) &=& 0
\\
B_1(\eta)\kappa_{n,1}(\lamp)+B_2(\eta)\kappa_{n,2}(\lamp) &=& 0
\label{green4}
\eea
Since $\kappa'_{n,1}(0)=0$ (see Appendix~D) and $\kappa'_{n,2}(0)\ne 0$, 
equation (\ref{green4a}) gives $A_2(\eta)=0$.
Combining equations (\ref{green1})--(\ref{green2}) and (\ref{green4}), 
we can express
\bea
A_1(\eta) &=& \frac{\kappa_{n,2}(\eta)\kappa_{n,1}(\lamp)-\kappa_{n,1}(\eta)\kappa_{n,2}(\lamp)}{D(\eta)}
\\
B_1(\eta) &=& -\frac{\kappa_{n,2}(\lamp)\kappa_{n,1}(\eta)}{D(\eta)}
\\
B_2(\eta) &=& \frac{\kappa_{n,1}(\lamp)\kappa_{n,1}(\eta)}{D(\eta)}
\eea
with the auxiliary function $D$ given by
\beq
D(\eta) = \left[\kappa_{n,1}(\eta)\kappa'_{n,2}(\eta)-\kappa'_{n,1}(\eta)\kappa_{n,2}(\eta)\right]\kappa_{n,1}(\lamp)
\eeq
and construct the Green function by substituting this back into (\ref{green0}).
The solution of (\ref{eqgp101}) is then given by
\bea\nonumber
\kapt(\xi) &=& \int_0^{\lamp} G(\xi,\eta)  \left(1-\phi\, \eee^{-{\eta^2}/{2\lambda_g^2}}  \right){\rm d}\eta = \\
&=&\int_0^{\lamp} G(\xi,\eta) \;{\rm d}\eta - \phi \int_0^{\lamp} G(\xi,\eta)\,\eee^{-{\eta^2}/{2\lambda_g^2}}\;{\rm d}\eta
\eea
This solution always satisfies conditions $\kapt'(0)=0$ and $\kapt(\lamp)=0$,
which have been incorporated into the Green function.
The remaining condition to be satisfied, $\kapt'(\lamp)=0$, 
provides a link between the load parameter and the size of the plastic zone.
As usual, it is more convenient to express the load parameter $\phi$ in terms
of the dimensionless plastic zone size $\lamp$ than vice versa.
Indeed, we can formally write
\beq
\kapt'(\xi) = \int_0^{\lamp} G'(\xi,\eta) \;{\rm d}\eta - \phi \int_0^{\lamp} G'(\xi,\eta)\,\eee^{-{\eta^2}/{2\lambda_g^2}}\;{\rm d}\eta
\eeq
where, for simplicity,  
\beq\label{green0x}
G'(\xi,\eta) = \frac{\partial G(\xi,\eta)}{\partial\xi}= \left\{ \begin{array}{ll}
A_1(\eta)\kappa'_{n,1}(\xi)
& \mbox{ for } 0\le\xi<\eta
\\
B_1(\eta)\kappa'_{n,1}(\xi)+B_2(\eta)\kappa'_{n,2}(\xi) & \mbox{ for } \eta<\xi<\lamp
\end{array}\right.
\eeq
denotes the partial derivative of the Green function
$G$ with respect to its first argument. From condition $\kapt'(\lamp)=0$
we obtain
\beq
\phi(\lamp,\lamg)=\frac{\int_0^{\lamp} G'(\lamp,\eta) \;{\rm d}\eta}{\int_0^{\lamp} G'(\lamp,\eta)\,\eee^{-{\eta^2}/{2\lambda_g^2}}\;{\rm d}\eta}
\eeq

\subsection{Results and Discussion}

\begin{figure}
\centering
\begin{tabular}{cc}
\hskip -20mm
(a) & (b)
\\
\hskip -20mm
\includegraphics[scale=0.6]{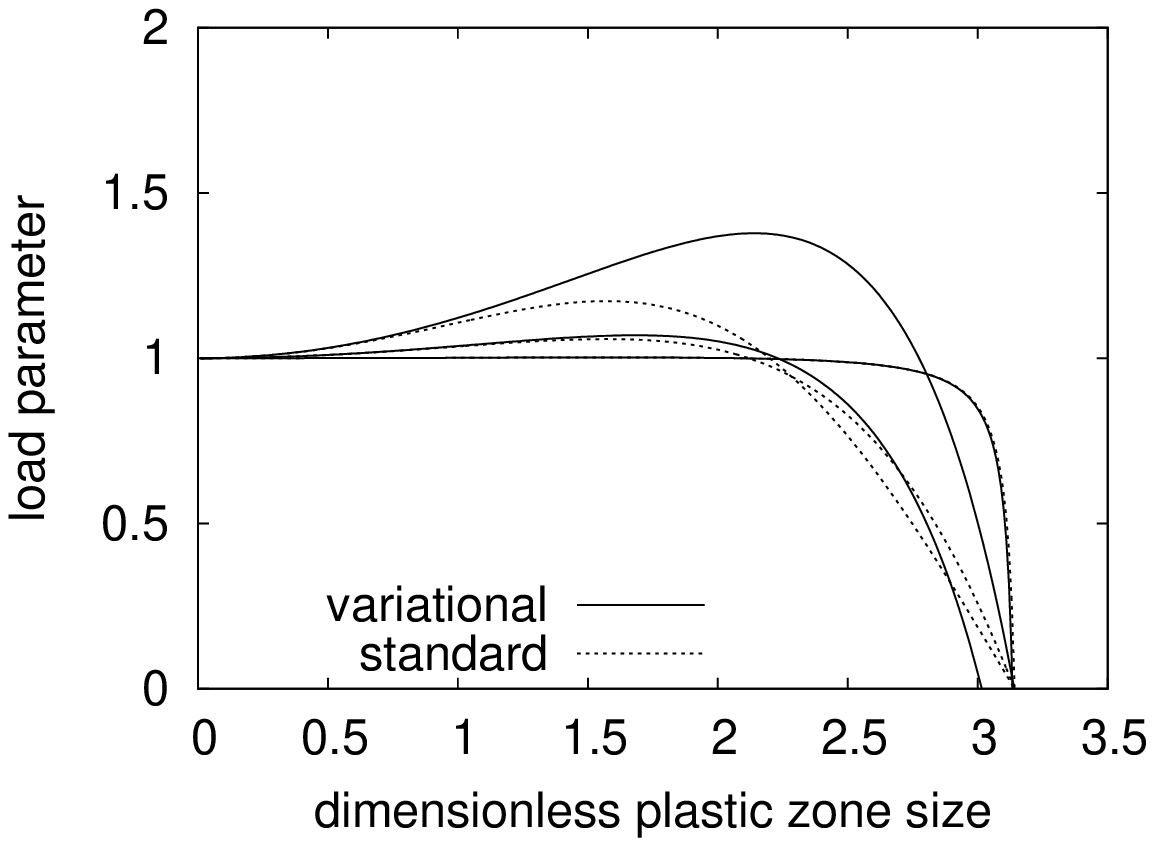}
&
\includegraphics[scale=0.6]{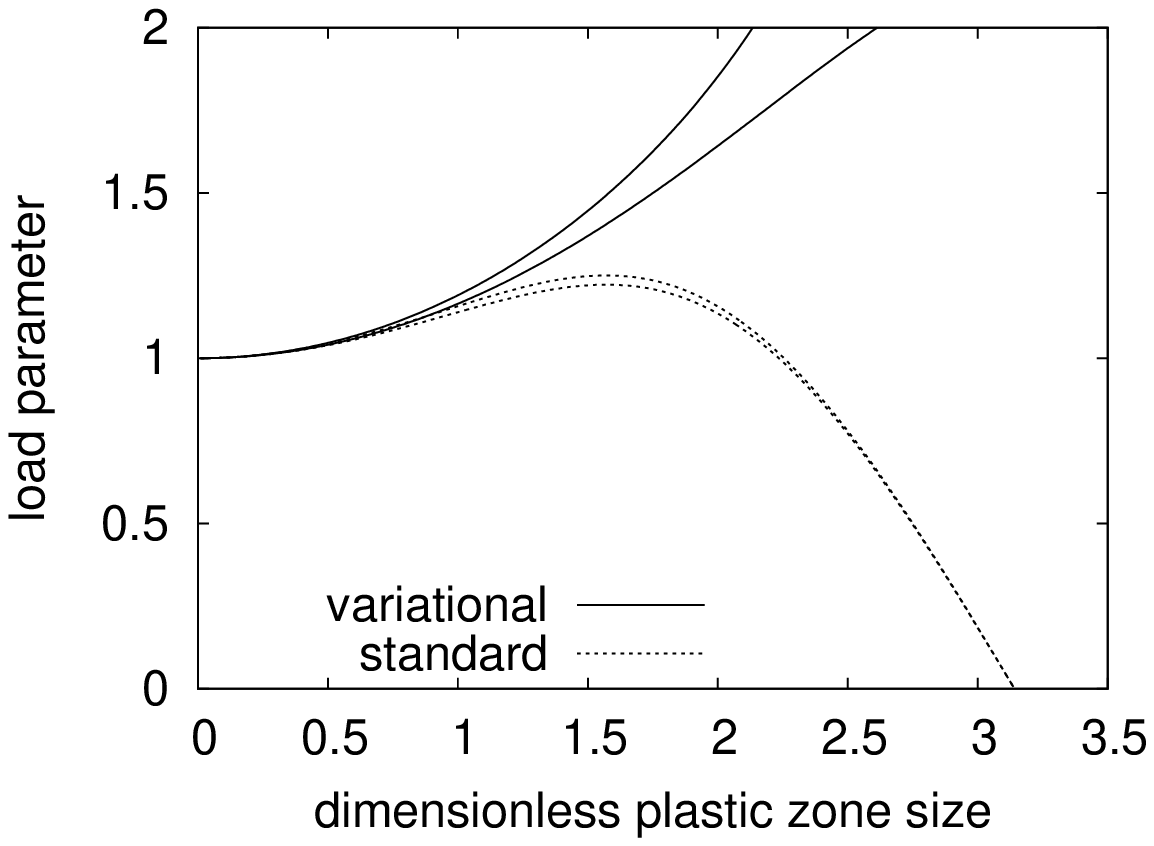}
\end{tabular}
\caption{Relation between load parameter and plastic zone size for a smooth
(Gaussian) stress distribution: (a) $\lamg=1.15$, 2, 10, (b) $\lamg=0.95$, 1.01
(from top to bottom)}
\label{fgp10}
\end{figure}

\begin{figure}
\centering
\begin{tabular}{cc}
\hskip -20mm
(a) & (b)
\\
\hskip -20mm
\includegraphics[scale=0.6]{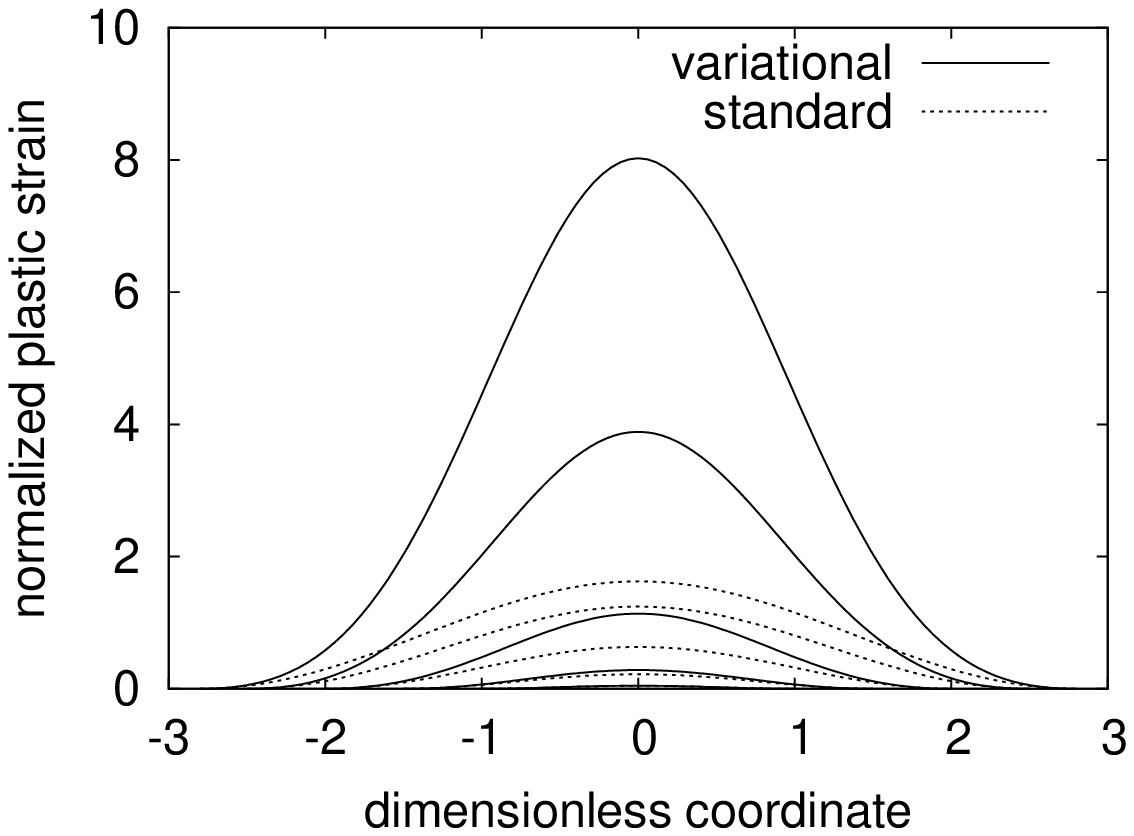}
&
\includegraphics[scale=0.6]{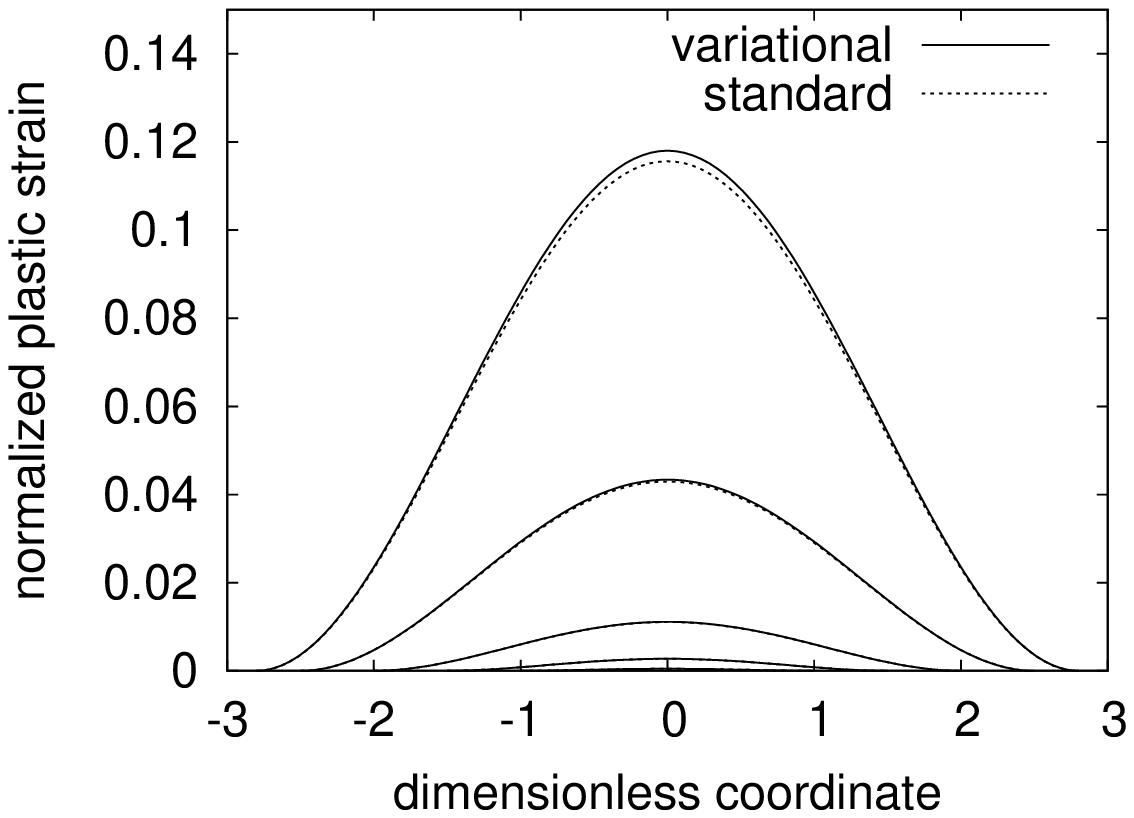}
\end{tabular}
\caption{Evolution of plastic strain profile for a smooth
(Gaussian) stress distribution: (a) $\lamg=1$, (b) $\lamg=10$}
\label{fgp8}
\end{figure}

\begin{figure}
\centering
\begin{tabular}{cc}
\hskip -20mm
(a) & (b)
\\
\hskip -20mm
\includegraphics[scale=0.6]{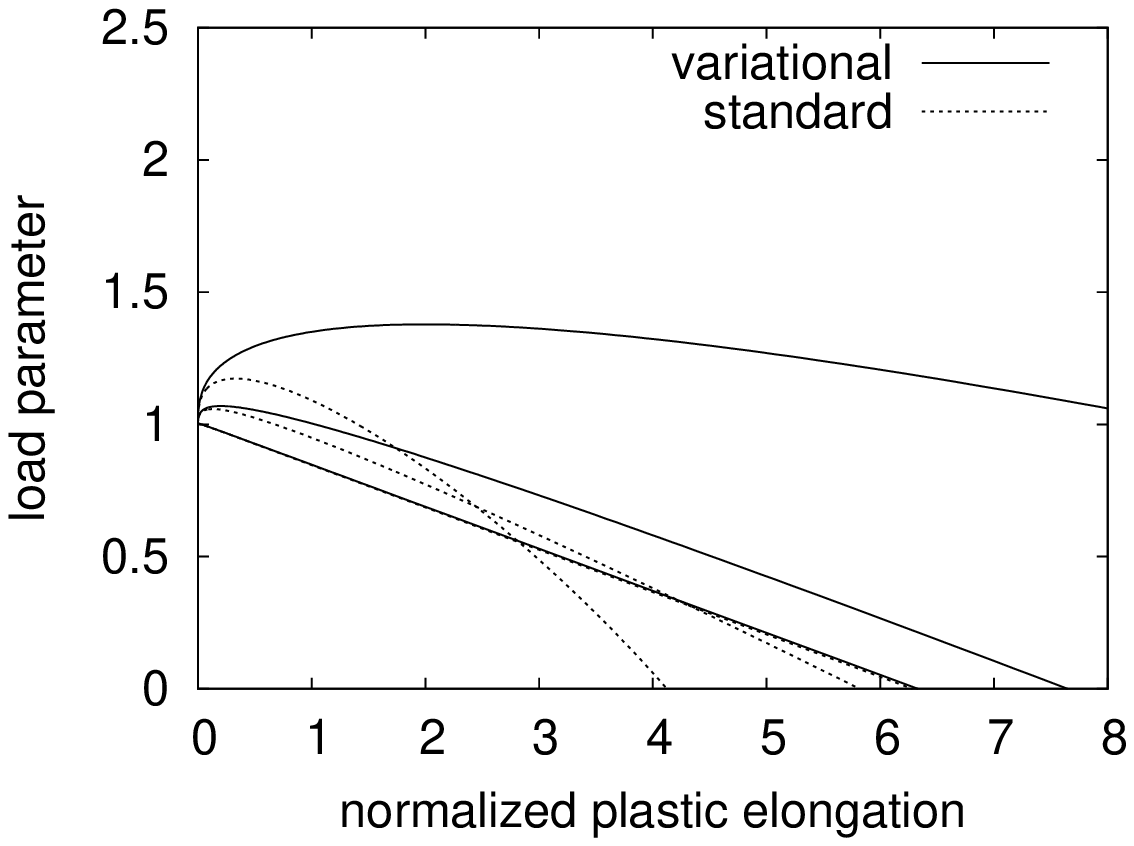}
&
\includegraphics[scale=0.6]{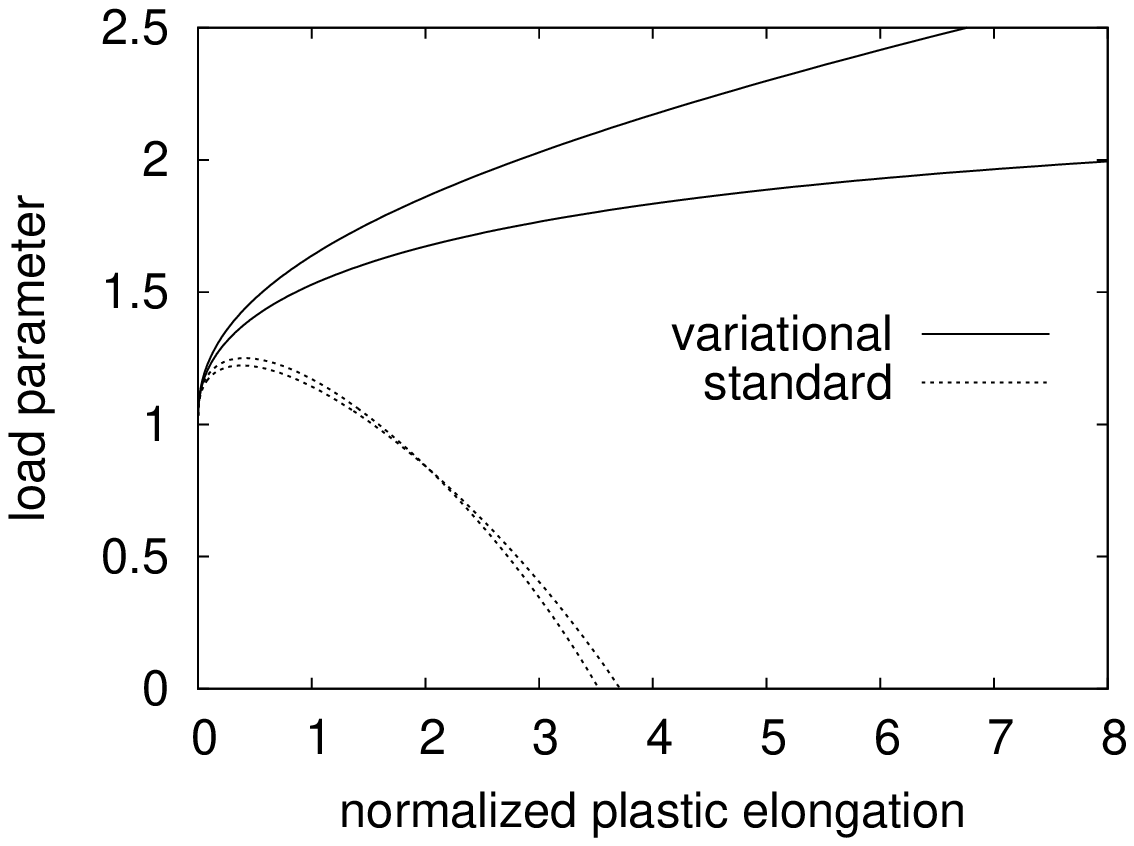}
\end{tabular}
\caption{Plastic part of normalized load-displacement diagram for a smooth
(Gaussian) stress distribution: (a) $\lamg=1.15$, 2, 10, (b) $\lamg=0.95$, 1.01 (from top to bottom)}
\label{fgp9}
\end{figure}

For illustration, the solution has been evaluated  and plotted
for a range of values of parameter $\lamg$. 
Fig.~\ref{fgp10} indicates that the plastic zone evolves continuously from the
weakest section and its size grows monotonically, which confirms
admissibility of the solution. 
Evolution of the plastic zone profile is shown in Fig.~\ref{fgp8}
for $\lamg=1$ and 10, and the plastic part of the normalized load-displacement
diagram is in Fig.~\ref{fgp9}.

In all these figures, solid curves correspond to the variational
formulation  from Section \ref{sec4.2}
and dotted curves to the standard one from Section \ref{sec4.1}.
For large values of $\lamg$, e.g.\ 10,
both formulations give almost identical results.
Initially, the plastic zone quickly expands 
and the load parameter $\phi$ increases only slightly from its value 1 at the
onset of plastic yielding. The softening part of the load-displacement
diagram is almost linear and as the load approaches zero, the plastic zone
size tends to  $2\pi l$ (for the standard formulation) or to a value
very close to $2\pi l$ (for the variational formulation). 
For intermediate values of $\lamg$,
e.g.\ 2 or 1.15, the hardening due to structural effects is more pronounced
for the variational formulation than for the standard one, and the 
load-displacement diagram becomes more ductile. Finally, for small 
values of $\lamg$, e.g.\ 1.01 or 0.95,  the variational formulation
gives unlimited hardening and unlimited expansion of the plastic zone,
while the standard formulation still leads to limited hardening
up to a finite peak load, followed by softening and expansion of the
plastic zone to its maximum size $2\pi l$.  

The strong hardening effect of the variational formulation for small
values of $\lamg$ can be attributed to the fact that the stabilizing
second term in (\ref{eq:50mod}) can be expanded into 
$Hl^2A(x)\kappa''(x)+Hl^2A'(x)\kappa'(x)$.
The second part, dependent on the derivative of the sectional area,
is neglected by the standard formulation but taken into account by the
variational one. For the problem considered here, the product $A'(x)\kappa'(x)$
is always negative (because $A(x)$ is decreasing in the left half of the
plastic zone and increasing in the right half, while  $\kappa(x)$ does the
opposite), and so the additional term
has a similar effect as negative $\kappa''(x)$. Such a term increases the
yield force and thus causes structural hardening accompanied by
an expansion of the plastic zone. 

It is important to note that cases  shown in
Fig.~\ref{fgp10}b and Fig.~\ref{fgp9}b 
correspond to an extreme nonuniformity of sectional area.
Indeed, $\lamg=1$ means that the sectional area at point $x=\pi l$,
which would be  the right boundary of the plastic zone,
is $\eee^{\pi^2/2}\approx 139$ times the area of the weakest section at the
center of the plastic zone. So this case is only of academic interest,
and is included here for completeness. 
\ignore
{
Parameter $\lamg$ has been defined as the ratio between the characteristic
length of the geometry, $\lg$, and the characteristic material length, $l$.
However, the specific 
physical meaning of $\lg$ is
 different for the three types of geometries considered in this
paper. For the bar with a weak segment, analyzed in Section \ref{},
$\lg$ is the half-length of the weak segment. In this case,
the ratio between the
sectional areas of the weak and standard sections is controled by another
parameter, $\beta$. For the bar with piecewise linear stress distribution,
 analyzed in Section \ref{}, $\lg$ denotes the 
}

\section{Summary and Conclusions}

In this paper, localization of plastic strain induced by a negative
plastic modulus has been studied using a variational formulation
of a gradient-enriched plasticity model. The main points can be
summarized as follows:
\begin{enumerate}
\item
Mathematical description of one-dimensional gradient plasticity
has been derived using a consistent variational approach, which naturally
provides not only the differential equation that represents 
the yield condition and needs to be satisfied inside the plastic 
zone, but  also the appropriate form of the boundary and jump conditions.
\item
Taking into account the jump conditions following from a variational principle,
a prototype problem with discontinuous data (in this specific case,
with discontinuous distribution of the sectional area) has been handled
successfully. It has been shown that the problem has a physically reasonable
solution, which can be constructed analytically.
\item
Two additional prototype problems, one with continuous but non-smooth data and 
the other with smooth data, have been analyzed, and analytical or 
semi-analytical solutions have been provided. The results have been 
compared to an alternative model that does not have a variational format.
\item
The influence of various parameters on the evolution of the
plastic zone, on the shape of the plastic strain profile and on the
resulting load-displacement diagram has been investigated for all three
cases mentioned above. It has been shown that the plastic zone evolves
from the weakest section and monotonically expands, which is in most cases
initially accompanied by an increase of the axial force over its elastic
limit value. This means that the structural response exhibits hardening
despite the softening character of the material model, which is related
to the expansion of the yielding process into stronger parts of the structure
induced by the gradient enrichment.
\item
The variational approach adopted here is based on the condition of non-negative
first variation of a certain energetic functional. States that satisfy 
this condition are considered as
valid solutions of the problem. It is expected that stable solutions 
correspond to local minima of the functional. A rigorous analysis
of the second variation, leading to an explicit stability criterion,
is presented in Appendix~A for the simplest case of a bar with uniform
properties.
\end{enumerate}

Finally, let us note that the approach elaborated here for the second-order
gradient model and for variable cross section can be extended to the
fourth-order gradient model and to variable material properties (such as
the yield stress or plastic modulus). The latter extension 
could be useful in studies
of the development of plastic zone near the interface of two materials
with different properties.

\appendix  
\section{Second Variation and Stability Conditions}

\newstuff
{
All analyses presented in the main body of the paper have been based
on the condition of non-negative first variation of functional $\Pi$.
Of course, this is only a necessary but not a sufficient condition for
a local minimum. Intuitively it can be expected that solutions corresponding
to a non-negative first variation but not to a local minimum are unstable.
Therefore, it is useful to investigate in more detail the changes of $\Pi$
in the immediate neighborhood of a ``point'' $(u,\kap)$ that represents
a valid solution, i.e., leads to a non-negative first variation $\delta\Pi$
for all admissible variations $\delta u$ and $\delta\kap$. 

If  $\delta\Pi>0$ for given  $\delta u$ and $\delta\kap$, the value of
$\Pi$ increases at least locally (for sufficiently small variations),
and the second variation does not need to be evaluated. However, for those
 $\delta u$ and $\delta\kap$ that give a vanishing first 
variation $\delta\Pi$, the sign of the increment is determined
by the second variation. So, as a first step, we restrict attention to those
combinations $\delta u$ and $\delta\kap$ for which  $\delta\Pi=0$.
Since the solution $(u,\kap)$ satisfies the equilibrium condition
(\ref{eqgp106x}), which is an equality, the first integral in (\ref{eqgp104}) vanishes
for an arbitrary displacement variation $\delta u$. On the other hand,
the yield condition (\ref{eq:50}) is an equality satisfied in the plastic
zone $\Ip$ only. Outside the plastic zone, the plastic admissibility
condition (\ref{eq:50x}) becomes an inequality. Leaving aside some degenerated
cases of neutral loading, this condition is typically satified as
a strict inequality. To make sure that the first variation $\delta\Pi$
vanishes, the plastic strain variation $\delta\kappa$ must be set to
zero outside the plastic zone. By similar arguments based on the jump
terms, it can be shown that the values of $\delta\kappa$ and of its derivative
on the boundary of the plastic zone must vanish. 

Since functional $\Pi$ given by (\ref{varfor1}) is quadratic,
its second variation is easily expressed as
\beq
\delta^2\Pi(\delta u,\delta\kap;u,\kap) = \intL \half EA(\delta u'-\delta\kap)^2\dx + \intL\half HA\left(\delta\kap^2-l^2\delta\kap'^2\right)\dx 
\label{varfor1x}
\eeq
Taking into account that $\delta\kap$ vanishes outside $\Ip$, as justified 
above, the integration domains can be reduced. After simple rearrangements
(with moduli $E$ and $H$ considered as constants), the resulting
stability condition can be written as
\beq
\int_{{\cal L}\setminus\Ip} A\delta u'^2\dx +\int_{\Ip} A(\delta u'-\delta\kap)^2\dx -\frac{H}{E} \int_{\Ip}A\left(l^2\delta\kap'^2-\delta\kap^2\right)\dx \ge 0
\label{varfor1y}
\eeq
In a loading test performed under displacement control, the displacements on the
physical boundary $\partial{\cal L}$ are prescribed and the variations
$\delta u$ vanish on $\partial{\cal L}$. 
Condition (\ref{varfor1y}) should be satisfied for all such $\delta u$ and
for all variations $\delta\kap$ with vanishing values and vanishing first
derivatives on the elasto-plastic boundary $\partial\Ip$. 

General analysis of condition (\ref{varfor1y})
for variable sectional area $A$ would be very tedious. However, the case
of a uniform bar (with $A=$ const.) is manageable, because (\ref{varfor1y})
simplifies to
\beq
\int_{{\cal L}\setminus\Ip} \delta u'^2\dx +\int_{\Ip} (\delta u'-\delta\kap)^2\dx -\frac{H}{E} \int_{\Ip}\left(l^2\delta\kap'^2-\delta\kap^2\right)\dx \ge 0
\label{varfor1z}
\eeq
In this case, the plastic zone $\Ip$ is an interval of length $\Lp=2\pi l$
(see Section~\ref{sec2.1}),
and the elastically unloading zone ${\cal L}\setminus\Ip$ is a union
of two intervals of total length $L-\Lp$.
Individual integrals in (\ref{varfor1z}) can be estimated from below.
For the second integral, we can exploit the Cauchy-Schwarz inequality,
which implies that
\beq\label{A4}
\int_{\Ip} (\delta u'-\delta\kap)^2\dx \ge \frac{1}{\Lp} \left(\int_{\Ip} (\delta u'-\delta\kap)\dx\right)^2=\Lp(\delta\bar{\eps}-\delta\bar{\kap})^2
\eeq
where
\beq
\delta\bar{\eps} = \frac{1}{\Lp}\int_{\Ip} \delta u'\dx, 
\hskip 10mm
\delta\bar{\kap} = \frac{1}{\Lp}\int_{\Ip} \delta\kap\dx
\eeq
are constants that represent the mean values of $\delta u'$ 
and $\delta\kap$ over the
plastic zone.  
The first integral in (\ref{varfor1z}) can be estimated in a similar fashion,  
taking into account that the mean value of $\delta u'$ in ${\cal L}\setminus\Ip$
is directly related to $\delta\bar{\eps}$, because of the compatibility constraint
$\intL \delta u'\dx = 0$:
\bea
&&\int_{{\cal L}\setminus\Ip} \delta u'\dx = - \int_{\Ip} \delta u'\dx  = - \Lp\delta\bar{\eps}\\
&&
\int_{{\cal L}\setminus\Ip} \delta u'^2\dx\ge \frac{1}{L-\Lp}\left(\int_{{\cal L}\setminus\Ip} \delta u'\dx\right)^2 = \frac{\Lp^2}{L-\Lp}\delta\bar{\eps}^2
\label{A7}
\eea
Finally, the last integral in (\ref{varfor1z}) can be estimated using the Wirtinger inequality, 

which is a one-dimensional case of the Poincar\'{e} inequality, 
with explicitly known
optimal constant. The theorem is applicable to the zero-mean part of 
$\delta\kap$ and implies that
\beq
\left(\frac{\Lp}{2\pi}\right)^2\int_{\Ip}\delta\kap'^2\dx \ge  \int_{\Ip}(\delta\kap-\delta\bar{\kap})^2\dx=\int_{\Ip}\delta\kap^2\dx -\Lp\delta\bar{\kap}^2
\eeq
Since $\Lp/2\pi= l$, we get
\beq\label{A9}
\int_{\Ip}\left(l^2\delta\kap'^2-\delta\kap^2\right)\dx \ge 
-\Lp\delta\bar{\kap}^2
\eeq

Substituting (\ref{A4}), (\ref{A7}) and (\ref{A9}) into (\ref{varfor1z}),
we obtain condition
\beq
\frac{\Lp^2}{L-\Lp}\delta\bar{\eps}^2 + \Lp(\delta\bar{\eps}-\delta\bar{\kap})^2+\frac{H}{E} \Lp\delta\bar{\kap}^2 \ge 0
\eeq
which means that a certain quadratic form of variables
$\delta\bar{\eps}$ and $\delta\bar{\kap}$ should be positive semidefinite. 
This leads to the following restrictions on the parameters:
\bea
\frac{\Lp^2}{L-\Lp} + \Lp \ge 0 & \Rightarrow & L\ge \Lp \\
\Lp + \frac{H}{E} \Lp\ge 0 & \Rightarrow & E+H\ge 0 \\
\left(\frac{\Lp^2}{L-\Lp} + \Lp\right)\left(\Lp + \frac{H}{E} \Lp\right)-\Lp^2 \ge 0 & \Rightarrow & 
\frac{\Lp}{L} \ge -\frac{H}{E}
\label{stabcond}
\eea
The first restriction corresponds to our tacit assumption that bar is longer
than $\Lp$, so that the full plastic zone can develop (for shorter bars,
the analysis would have to be modified). The second restriction excludes
snapback of the stress-strain diagram with no gradient effects.
The third restriction is the most stringent one.
It guarantees stability of the localized solution under displacement control
and can be interpreted e.g.\ as a constraint on the maximum length of the
bar (with respect to the characteristic length $l$, reflected by the plastic
zone size $\Lp=2\pi l$). It is reassuring that this condition exactly
coincides with the condition of negative slope of the post-peak 
load-displacement diagram. Indeed, the total elongation of the bar
after the onset of plastic yielding can be expressed as a sum of the
elastic and plastic parts:
\beq
u_{tot} = \intL\eps\dx = \intL \frac{F}{EA}\dx + \int_{\Ip}\kap\dx =
\frac{FL}{EA} + u_p
\eeq
Substituting the plastic strain distribution according to (\ref{varfor12}),
which refers to the plastic zone $\Ip=(-\pi l,\pi l)$, the
plastic part of elongation turns out to be
\beq
u_p = \int_{-\pi l}^{\pi l} \frac{F/A-\sigma_0}{H}\left(1+\cos\frac{x}{l}\right)\dx = \frac{F/A-\sigma_0}{H}2\pi l = \frac{F\Lp}{HA} - \frac{\sigma_0 \Lp}{H}
\eeq
The slope of the post-peak part of load-displacement diagram is
the reciprocal value of the tangent structural compliance $L/EA + \Lp/HA$. 
Snapback occurs if the tangent compliance (and thus also the tangent stiffness) 
is positive, i.e., if
\beq
\frac{L}{E}+\frac{\Lp}{H} > 0
\eeq
This is of course equivalent to $L/\Lp<-H/E$, which holds if and only if
the stability condition (\ref{stabcond}) is violated.

To illustrate the difference between the condition of non-negative
first variation and the (stronger) condition of a local minimum,
let us recall that the localization problem for a uniform bar admits
not only solutions with the plastic zone of size $\Lp=2\pi l$, but
also solutions with $\Lp$ equal to integer multiples of $2\pi l$;
see Section~\ref{sec2.1}. 
\ignore{
To give a specific example, 
suppose that a bar of length $L=5\pi l$
is represented by the interval $[-2.5\pi l, 2.5\pi l]$, and that the 
plastic modulus $H=-0.39E$. Solutions with plastic strain localized into 
an interval of size $\Lp=2\pi l$ satisfy condition (\ref{stabcond}) and
thus are stable (under displacement control). However, there exist also
solutions with the plastic zone  of length $\Lp=4\pi l$. 
}
For instance, a plastic strain distribution
given by
\beql{varfor12z}
\kap(x) = \left\{
\begin{array}{ll}
\displaystyle\frac{\sig-\sig_0}{H} \left(1-\cos\frac{x}{l}\right) & \mbox{ for } x\in\Ip=(-2\pi l,2\pi l)\\
0 & \mbox{ for } x\notin\Ip
\end{array}\right.
\eeq 
satisfies differential equation (\ref{eq:50z}) as well as conditions 
$\kap=0$ and $\kap'=0$ at the boundary of $\Ip$, i.e., at points $\pm 2\pi l$.
The length of the plastic zone, $\Lp=4\pi l$, is now the double of the 
minimum plastic zone length. 
Consider an admissible variation of $\kap$ in the form
\beq
\delta\kap(x) = \left\{
\begin{array}{ll}
c\; \sgn(x)\displaystyle\left(1-\cos\frac{x}{l}\right) & \mbox{ for } x\in\Ip\\
0 & \mbox{ for } x\notin\Ip
\end{array}\right.
\eeq
where  $c$ is an arbitrary constant, not exceeding in magnitude the positive
constant $(\sig-\sig_0)/H$
(for larger $|c|$, the function $\kap+\delta\kap$ would not be non-negative and would not belong to the space of admissible functions $V_\kap$
defined in (\ref{varfor3})).  A simple calculation reveals that
\beq\label{a19}
\int_{\Ip}\left(l^2\delta\kap'^2-\delta\kap^2\right)\dx =
\int_{-2\pi l}^{2\pi l} c^2\left[\sin^2\frac{x}{l} - \left(1-\cos\frac{x}{l}\right)^2\right]\dx = -4c^2\pi l
\eeq
Since the mean value of $\delta\kap$ is zero, the displacement variation 
$\delta u$ can be selected such that $\delta u'=\delta\kap$, and then the
first two integrals in (\ref{varfor1z}) vanish. The third integral,
evaluated in (\ref{a19}), is negative and
is multiplied by a positive constant $-H/E$, and so the stability condition  
(\ref{varfor1z}) is
always violated. 
This proves that solutions with plastic zone exceeding
the minimum length $2\pi l$ would be unstable, 
independently of the total bar length $L$.
\ignore
{
combined with an admissible variation
of $u$ in the form 
\beq
\delta u(x) = \left\{
\begin{array}{ll}
0 & \mbox{ for } -2.5\pi l\le x \le -2\pi l \\
c \left(x+2\pi l\right) & \mbox{ for } x\in\Ip\\
1 & \mbox{ for } 2.5\pi l\le x \le 2.5\pi l
\end{array}\right.
\eeq
leads to a negative value of the expression on the left-hand side of (\ref{varfor1z}),
as can be easily verified:
\beq
\int_{{\cal L}\setminus\Ip} \delta u'^2\dx +\int_{\Ip} (\delta u'-\delta\kap)^2\dx -\frac{H}{E} \int_{\Ip}\left(l^2\delta\kap'^2-\delta\kap^2\right)\dx =
0+ ...
\eeq
A solution of this kind does not correspond to a local minimum of
functional $\Pi$ and is considered as unstable. 
}
}

\section{General Solution of Equation (\ref{eqgp3})}

For negative $\xi$, equation (\ref{eq-gp3}) can be written as
\begin{equation}
\kapt''(\xi) -\frac{1}{\lambda_g+\xi}\kapt'(\xi)+\kapt(\xi) =1-\phi
-\frac{\phi}{\lamg}\xi
\label{eqgp3x}
\end{equation}
Using substitutions
\bea
\xi &=& \eta-\lamg \\
\kappa(\xi)&=&(\lamg+\xi)\,g(\lamg+\xi)=\eta\,g(\eta)
\label{gbess}
\eea
where $g$ is a new unknown function and $\eta$ is a shifted dimensionless
spatial coordinate,  
we can convert (\ref{eqgp3x}) into 
\beql{gbess2}
\eta^2 g''(\eta) + \eta g'(\eta) + (\eta^2-1)g(\eta) = \eta - \frac{\phi}{\lamg}\eta^2
\eeq
where 
primes
denote derivatives with respect to $\eta$. 
The homogeneous counterpart of (\ref{gbess2}) is
the Bessel equation of order $\nu=1$, and its general solution
can be written as
\begin{equation}
g_h(\eta)=C_1\mbox{J}_1(\eta)+C_2\mbox{Y}_1(\eta)
\label{eqgp9}
\end{equation}
where $C_1$ and $C_2$ are arbitrary constants,
and $\mbox{J}_1$ and $\mbox{Y}_1$ are respectively the Bessel functions 
of the first and second kind; see \cite{uvodbessel}. 

Now we need to find a particular solution for the given 
right-hand side. It turns out that the expression on the left-hand side 
of (\ref{gbess2})
gives a multiple of $\eta$ if $g$ is set simply to $1/\eta$, and a multiple
of $\eta^2$ if $g$ is set to the Struve function $\mbox{\bf H}_1(x)$; 
see \cite{uvodbessel}. Therefore, we look for the particular solution
in the form
\beql{gbess4}
\tilde{g}(\eta) = \frac{k_1}{\eta} + k_2  \mbox{\bf H}_1(\eta)
\eeq
and substituting into the left-hand side of (\ref{gbess2}) we get
the condition
\beq
k_1\eta + \frac{2k_2\eta^2}{\pi} = \eta - \frac{\phi}{\lamg}\eta^2
\eeq
from which $k_1=1$ and $k_2=-\pi\phi/2\lamg$. Combining the particular
solution $\tilde{g}$ given by (\ref{gbess4}) 
with the general solution of the homogeneous equation $g_h$ given 
 by (\ref{eqgp9}) and substituting this back into  (\ref{gbess}), 
we get the general solution of  (\ref{eqgp3x}) in the form
\beq
\kappa(\xi) = (\lamg+\xi)[C_1\mbox{J}_1(\lamg+\xi)+C_2\mbox{Y}_1(\lamg+\xi)]
+1 - \frac{\pi\phi}{2\lamg}(\lamg+\xi)\mbox{\bf H}_1(\lamg+\xi)
\eeq
Recall that  (\ref{eqgp3x}) is the specific form of (\ref{eq-gp3})
valid for $\xi<0$. For $\xi>0$, it is sufficient to replace $\lamg+\xi$ by
$\lamg-\xi$. The resulting expression valid for both positive and negative
$\xi$ is given in (\ref{eqgp20}). 

\section{General Solution of Equation (\ref{eqgp125})} 

The integrals in (\ref{eqc1})--(\ref{eqc2}) are most conveniently evaluated
if one introduces an 
auxiliary complex function
\beq
\tilde{C}(\xi)=\tilde{C}_1(\xi)-i\tilde{C}_2(\xi)
\eeq
Substituting from (\ref{eqc1})--(\ref{eqc2}), one gets
\bea\nonumber
\tilde{C}(\xi)&=&\cos\xi-i\sin\xi +\phi 
\int\eee^{-{\xi^2}/{2\lambda_g^2}}(\sin\xi+i\cos\xi) \dxi
=\\
\label{eq100}
&=&\eee^{-i\xi} +i\phi \int\eee^{-{\xi^2}/{2\lambda_g^2}}\eee^{-i\xi} \dxi
\eea
Since
\beq
-\frac{\xi^2}{2\lamg^2}-i\xi = -\frac{(\xi+i\lamg^2)^2+\lamg^4}{2\lamg^2}=
-\frac{(\xi+i\lamg^2)^2}{2\lamg^2}-\frac{\lamg^2}{2}
\eeq
the last integral in (\ref{eq100}) can be written as
\beql{eq102}
\int\eee^{-{\xi^2}/{2\lambda_g^2}}\eee^{-i\xi} \dxi=
\eee^{-\lamg^2/2}\int \exp\left(-\frac{(\xi+i\lamg^2)^2}{2\lamg^2}\right)\dxi
\eeq
and substitution $\xi=\sqrt{2}\lamg t-i\lamg^2$ leads to
\bea\nonumber
\int \exp\left(-\frac{(\xi+i\lamg^2)^2}{2\lamg^2}\right)\dxi&=&
\sqrt{2}\lamg\int\eee^{-t^2}\mbox{d}t = 
\lamg\sqrt{\frac{\pi}{2}}\erf(t)=\\
\label{eq103}
&=&
\lamg\sqrt{\frac{\pi}{2}}\erf\left(\frac{\xi+i\lamg^2}{\sqrt{2}\lamg}\right)
\eea
where $\erf$ is the ``error function'' defined by
\beq
\erf(x) = \frac{2}{\sqrt{\pi}}\int\eee^{-x^2}\dx, \hskip 10mm \erf(0)=0
\eeq
Combining (\ref{eq102}) and (\ref{eq103}) and substituting 
back into (\ref{eq100}), we obtain
\beql{eq105}
\tilde{C}(\xi)=
\eee^{-i\xi} +i\phi \lamg\sqrt{\frac{\pi}{2}}\eee^{-\lamg^2/2}\erf\left(\frac{\xi+i\lamg^2}{\sqrt{2}\lamg}\right)
\eeq
Functions $\tilde{C}_1$ and $\tilde{C}_2$ could now be extracted from the
real and imaginary part of the expression in (\ref{eq105}). But this is
not even necessary, because the particular solution of equation 
(\ref{eqgp125}) given by
formula  (\ref{eqgp125y}) can be presented as
\beq
\tilde{\kappa}_{\rm n}(\xi)=\mbox{Re}\left[
(\tilde{C}_1(\xi)-i\tilde{C}_2(\xi))(\cos\xi+i\sin\xi)\right]=
\mbox{Re}\left[\tilde{C}(\xi)\eee^{i\xi}\right]
\eeq
where $\mbox{Re}$ stands for the real part.
Substituting for $\tilde{C}(\xi)$ according to (\ref{eq105}),
we get the particular solution in the form
\beq
\tilde{\kappa}_{\rm n}(\xi)=1+\phi \lamg\sqrt{\frac{\pi}{2}}\eee^{-\lamg^2/2}\mbox{Re}\left[i\,\erf\left(\frac{\xi+i\lamg^2}{\sqrt{2}\lamg}\right)\eee^{i\xi}\right]
\eeq
Recalling the definition of the Dawson function \cite{olver},
\begin{equation}
\mbox{F}(x)=e^{-x^2}\int\limits_0^x e^{t^2}\mbox{d}t=-\frac{i\sqrt{\pi}}{2}e^{-x^2}\erf(ix)
\label{eq201}
\end{equation}
and taking into account that $\overline{F(x)}=F(\bar{x})$ (with the bar denoting
the complex conjugate), we can rewrite the result as
\beq
\tilde{\kappa}_{\rm n}(\xi)=1-\phi \lamg\eee^{-\xi^2/2\lamg^2}\frac{\sqrt{2}}{2}
\left[F\left(\frac{\lamg^2-i\xi}{\sqrt{2}\lamg}\right)+F\left(\frac{\lamg^2+i\xi}{\sqrt{2}\lamg}\right)\right]
\eeq
The general solution (\ref{eqdaws}) 
of equation (\ref{eqgp125}) is then obtained by adding
a linear combination of functions $\cos\xi$ and $\sin\xi$ with arbitrary
coefficients.

\section{General Solution of Homogeneous Form of Equation~(\ref{eqgp101})} 

The homogeneous counterpart of equation (\ref{eqgp101}) can be written as
\begin{equation}
\kappa_n''(\xi)+\frac{\xi}{\lambda_g^2}\kappa_n'(\xi)+\kappa_n(\xi)=0
\label{eq3}
\end{equation}
Defining a rescaled spatial variable $\eta=\xi/\lamg\sqrt{2}$ and 
expressing the unknown function as
\beq\label{rescaled}
\kappa_n(\xi)=\eee^{-\frac{\xi^2}{2\lambda_g^2}}u\left(\frac{\xi}{\lamg\sqrt{2}}\right)=
\eee^{-\eta^2} u(\eta)
\eeq
where $u$ is a transformed unknown function, we can convert (\ref{eq3}) 
into the so-called Hermite differential equation,
\begin{equation}
u''(\eta)-2\eta u'(\eta)+2\nu u(\eta)=0
\label{eq5}
\end{equation}
where $\nu=\lambda_g^2-1$ is a real parameter 
(in our case larger than $-1$)
and primes denote differentiation with
respect to $\eta$. 

The solution of the Hermite equation (\ref{eq5}) can be constructed in terms
of infinite power series
\begin{equation}
u(\eta)=\sum_{r=0}^{\infty}{a_r\eta^{r}}
\label{eq4}
\end{equation}
Expressing the derivatives
\bea
u'(\eta)&=&\sum_{r=1}^{\infty}{a_{r}r\eta^{r-1}}\\
u''(\eta)&=&\sum_{s=2}^{\infty}{a_{s}s(s-1)\eta^{s-2}}
\eea
and substituting them into  (\ref{eq5}), we obtain
\beq
\sum_{s=2}^{\infty}{a_{s}s(s-1)\eta^{s-2}}-2 \sum_{r=1}^{\infty}{a_{r}r\eta^{r}}
+2\nu \sum_{r=0}^{\infty}{a_r\eta^{r}} = 0
\eeq
In the first sum, $s$ can be replaced by $r+2$, with $r$
running from 0 to infinity, and in the second sum, $r$ can also run from 0
without changing the result (because of the presence of the factor $r$
which anihilates the term with $r=0$). The equation can thus be rewritten
as
\beq
\sum_{r=0}^{\infty}\left[a_{r+2}(r+2)(r+1)-2 a_{r}r +2\nu a_r  \right]    \eta^{r} = 0
\eeq
and the term in the square brackets must vanish for each individual value
of $r=0,1,2,\ldots$. This condition results into the recursive formula
\begin{equation}
a_{r+2}=\frac{2(r-\nu)}{(r+1)(r+2)}a_r, \hskip 10mm r=0,1,2,\ldots
\label{eq7}
\end{equation}
Note that $a_{r+2}$ is expressed in terms of $a_r$. It is therefore possible
to select arbitrary values of $a_0$ and $a_1$, and then express
all other coefficients
with even subscripts in terms of $a_0$ and all 
other coefficients
with odd subscripts in terms of $a_1$. 
Setting $a_0=1$ and $a_1=0$, or $a_0=0$ and $a_1=1$,
leads to
two linearly independent functions
\bea\nonumber
u_1(\eta)&=&1-\frac{2\nu}{2!}\eta^2+\frac{2^2\nu(\nu-2)}{4!}\eta^4-\frac{2^3\nu(\nu-2)(\nu-4)}{6!}\eta^6+\ldots
\\
\label{eq8}
\\
u_2(\eta)&=&\eta-\frac{2(\nu-1)}{3!}\eta^3+\frac{2^2(\nu-1)(\nu-3)}{5!}\eta^5+\ldots
\label{eq10}
\eea
and every solution of equation (\ref{eq5}) can be expressed as their
linear combination.
The fundamental solutions 
can be conveniently expressed in terms of the so-called
confluent hypergeometric function of the first kind,
denoted as $_{1}\hskip -0.5mm F_1(\alpha;\gamma;x)$ (note that the symbol $F$ has a left subscript and a right subscript), which is defined by the infinite series \cite{Sneddon}
\begin{equation}
_{1}\hskip -0.5mm F_1(\alpha;\gamma;x)=\sum_{r=0}^{\infty}\frac{(\alpha)_r}{(\gamma)_r}\frac{x^r}{r!}
\label{eq11}
\end{equation}
Here, $(\bullet)_r$ is the so-called Pochhammer symbol,
which can be expressed in terms of  Euler's gamma function:
\begin{equation}
(\alpha)_r=\alpha(\alpha+1)\dots(\alpha+r-1)=\frac{\Gamma(\alpha+r)}{\Gamma(\alpha)}
\label{eq12}
\end{equation}
It is easy to verify by simple substitution that the fundamental solutions $u_1$ and $u_2$
from (\ref{eq8})--(\ref{eq10}) can be rewritten as
\bea
u_1(\eta)&=& _1\hskip -0.5mm F_1(-\nu/2;1/2;\eta^2)
\label{eq13}
\\
u_2(\eta)&=&\eta\; _{1}\hskip -0.5mm F_1((1-\nu)/2;3/2;\eta^2)
\label{eq14}
\eea
Substituting this into (\ref{rescaled})
and replacing $\nu$ by $\lambda_g^2-1$ and $\eta$ by 
$\xi/\lamg\sqrt{2}$, we finally obtain two linearly independent
solutions of (\ref{eq3}) in the form
\bea
\kappa_{n,1}(\xi)&=&\eee^{-\frac{\xi^2}{2\lambda_g^2}}\; {_1}\hskip -0.5mm \mbox{F}_1\left(\frac{1-\lambda_g^2}{2};\frac{1}{2};\frac{\xi^2}{2\lambda_g^2}\right)
\label{eq15}
\\
\kappa_{n,2}(\xi)&=&\eee^{-\frac{\xi^2}{2\lambda_g^2}}\frac{\xi}{\sqrt{2}\lambda_g} {_1}\hskip -0.5mm \mbox{F}_1\left(\frac{2-\lambda_g^2}{2};\frac{3}{2};\frac{\xi^2}{2\lambda_g^2}\right)
\label{eq15x}
\eea
Note that function $u_1$ is even and $u_2$ is odd, and so $\kappa_{n,1}$ is even
and $\kappa_{n,2}$ is odd. One useful consequence is that $\kappa_{n,1}'(0)=0$.

\ignore
{
Further, substitution $\zeta=\eta^2$ and $u(\eta)=v(\eta^2)=v(\zeta)$ leads 
to the so-called confluent hypergeometric differential equation
\begin{equation}
\zeta v''(\zeta)+\left(\frac{1}{2}-\zeta\right)v'(\zeta)+\frac{\nu}{2}v(\zeta)=0
\label{eq6}
\end{equation}
which can be solved in terms of confluent hypergeometric functions and Hermite polynomials \cite{Sneddon}. The general solution has the form
\bea\nonumber
v(\zeta)&=&A\; {_1}\mbox{F}_1\left(-\frac{\nu}{2},\frac{1}{2};\zeta\right)+B\sqrt{\zeta}\; {_1}\mbox{F}_1\left(-\frac{1}{2}(\nu-1),\frac{3}{2};\zeta\right)=
\\
&=&A\; {_1}\mbox{F}_1\left(-\frac{\nu}{2},\frac{1}{2};\zeta\right)+C\mbox{H}_\nu\left(\sqrt{\zeta}\right)
\label{eq7}
\eea
where $A$, $B$ and $C$ are constants. 
Symbols ${_1}\mbox{F}_1$ and $\mbox{H}_\nu$ denote the confluent hypergeometric function of the first kind \cite{Sneddon} and a Hermite polynomial ??????????????
Finally, back substitution into (\ref{rescaled}) gives
the general solution of the homogeneous version of equation (\ref{eqgp101})
in the form
\begin{equation}
\kappa_n(\xi)=A \eee^{-\frac{\xi^2}{2\lambda_g^2}} {_1}\mbox{F}_1\left(\frac{1}{2}(1-\lambda_g^2),\frac{1}{2};\frac{\xi^2}{2\lambda_g^2}\right)+C \eee^{-\frac{\xi^2}{2\lambda_g^2}}\mbox{H}_{(\lambda_g^2-1)}\left(\frac{\xi}{\lambda_g\sqrt{2}}\right)
\label{eq8}
\end{equation}

} 


\section*{Acknowledgements}
Financial support of the Czech Science Foundation (GA\v{C}R)
under projects 106/08/1508, 201/10/0357 and 108/11/1243 
is gratefully acknowledged. The authors would also like to thank
Dr.~Milada Kop\'{a}\v{c}kov\'{a} (Institute of Mathematics,
Academy of Sciences of the Czech Republic) 
and Mr.~Jaroslav Vond\v{r}ejc (PhD student
at the Czech Technical University) for stimulating discussions.



\begin{thebibliography}{00}
\bibitem[\protect\citeauthoryear{Aifantis}{1984}]{Aif84}
E. C. Aifantis.
\newblock {A gradient flow theory of plasticity for granular materials}.
\newblock {\em Archives of Mechanics},
87: 197--217, 1984.
%
\bibitem[\protect\citeauthoryear{Andrews}{1992}]{2222} 
L. C. Andrews.
Special functions of mathematics for engineers,  McGraw-Hill, Inc.
%
\bibitem[\protect\citeauthoryear{Borino, Fuschi and Polizzotto}{1999}]{BorFusPol99}
G. Borino, P. Fuschi and C. Polizzotto.
\newblock {A thermodynamic approach to nonlocal plasticity and
		  related variational approaches}.
\newblock {\em Journal of Applied Mechanics},
66: 952--963, 1999.
%
\bibitem[\protect\citeauthoryear{de Borst and M\"{u}hlhaus}{1992}]{bormuhl92}
R. de Borst and H. B. M\"{u}hlhaus.
\newblock {Gradient-dependent plasticity: Formulation and algorithmic aspects}.
\newblock {\em International Journal for Numerical Methods in Engineering},
35:521--539, 1992.
%
\bibitem[\protect\citeauthoryear{Jir\'{a}sek and Ba\v{z}ant}{2001}]{JirBaz01}
M. Jir\'{a}sek and Z. P. Ba\v{z}ant.
\newblock {\em Inelastic Analysis of Structures}.
\newblock John Wiley and Sons, 2001.
%
\bibitem[\protect\citeauthoryear{Korenev}{2002}]{uvodbessel} 
B. G. Korenev, 
\newblock {\em Bessel Functions and their Applications}. 
\newblock Taylor \& Francis, 2002.
\bibitem[\protect\citeauthoryear{Jir\'{a}sek, Zeman and Vond\v{r}ejc}{2010}]{jzv10} 
M. Jir\'{a}sek, J. Zeman and J. Vond\v{r}ejc.
\newblock {Softening gradient plasticity: Analytical study of localization under nonuniform stress}.
\newblock {\em International Journal for Multiscale Computational Engineering}, 8: 37--60, 2010.
%
\bibitem[\protect\citeauthoryear{Liebe and Steinmann}{2001}]{liestei01}
T. Liebe and P. Steinmann.
\newblock {Theory and numerics of a thermodynamically consistent framework for geometrically linear gradient plasticity}.
\newblock {\em International Journal for Numerical Methods in Engineering},
51: 1437-1467, 2001.
%
\bibitem[\protect\citeauthoryear{Lubliner}{1990}]{Lub90}
J. Lubliner.
\newblock {\em Plasticity Theory}.
\newblock Macmillan Publishing Company, 1990.
%
\bibitem[\protect\citeauthoryear{M\"{u}hlhaus and Aifantis}{1991}]{muhlaif91}
H. B. M\"{u}hlhaus and E. C. Aifantis.
\newblock {A variational principle for gradient plasticity}.
\newblock {\em International Journal of Solids and Structures},
28:845--858, 1991.
%
\bibitem[\protect\citeauthoryear{Olver}{1997}]{olver} 
F. W. J. Olver.
\newblock {\em Asymptotics and special functions}. 
\newblock A K Peters, Ltd., 1997.
%
\bibitem[\protect\citeauthoryear{Petryk}{2003}]{Pet03}
H. Petryk.
\newblock {Incremental energy minimization in dissipative solids}.
\newblock {\em Comptes Rendus M\'{e}canique},
331: 469--474, 2003.
%
\bibitem[\protect\citeauthoryear{Polizzotto, Borino and Fuschi}{1998}]{PolBorFus98}
C. Polizzotto, G. Borino and P. Fuschi.
\newblock {A thermodynamic consistent formulation of nonlocal
		  and gradient plasticity}.
\newblock {\em Mechanics Research Communications},
25: 75--82, 1998.
%
\bibitem[\protect\citeauthoryear{Sneddon}{1956}]{Sneddon} 
I. N. Sneddon.
\newblock {\em Special functions of mathematical physics and chemistry}. 
\newblock Oliver and Boyd, 1956.
%
\bibitem[\protect\citeauthoryear{Svedberg}{1996}]{sved96}
T. Svedberg.
\newblock {\em A Thermodynamically Consistent Theory of Gradient-Regularized Plasticity Coupled to Damage}.
\newblock Licentiate Thesis, Chalmers University of Technology, 1996.
%
\bibitem[\protect\citeauthoryear{Svedberg and Runesson}{1997}]{SveRun97}
T. Svedberg and K. Runesson.
\newblock {A thermodynamically consistent theory of gradient-regularized plasticity coupled to damage}.
\newblock {International Journal of Plasticity},
13: 669--696, 1997.
%
\bibitem[\protect\citeauthoryear{Svedberg and Runesson}{1998}]{SveRun98a}
T. Svedberg and K. Runesson.
\newblock {Thermodynamically consistent nonlocal and gradient formulations of plasticity}.
\newblock {\em Nonlocal Aspects in Solid Mechanics}, EUROMECH Colloquium 378,
Mulhouse, France, 32--37, 1998.
%
\bibitem[\protect\citeauthoryear{Valanis}{1996}]{Val96}
K. C. Valanis.
\newblock {A gradient theory of internal variables}.
\newblock {\em Acta Mechanica},
116: 1--14, 1996.
\end{thebibliography}



\end{document}